\documentclass{jfm}

\ifCUPmtlplainloaded \else
  \checkfont{eurm10}
  \iffontfound
    \IfFileExists{upmath.sty}
      {\typeout{^^JFound AMS Euler Roman fonts on the system,
                   using the 'upmath' package.^^J}%
       \usepackage{upmath}}
      {\typeout{^^JFound AMS Euler Roman fonts on the system, but you
                   dont seem to have the}%
       \typeout{'upmath' package installed. JFM.cls can take advantage
                 of these fonts,^^Jif you use 'upmath' package.^^J}%
       \providecommand\upi{\pi}%
      }
  \else
    \providecommand\upi{\pi}%
  \fi
\fi

\ifCUPmtlplainloaded \else
  \checkfont{msam10}
  \iffontfound
    \IfFileExists{amssymb.sty}
      {\typeout{^^JFound AMS Symbol fonts on the system, using the
                'amssymb' package.^^J}%
       \usepackage{amssymb}%
         \let\leq=\leqslant
       \let\ge=\geqslant  \let\geq=\geqslant
      }{}
  \fi
\fi

\ifCUPmtlplainloaded \else
  \IfFileExists{amsbsy.sty}
    {\typeout{^^JFound the 'amsbsy' package on the system, using it.^^J}%
     \usepackage{amsbsy}}
    {\providecommand\boldsymbol[1]{\mbox{\boldmath $##1$}}}
\fi

\providecommand\bnabla{\boldsymbol{\nabla}}

\newsavebox{\astrutbox}
\sbox{\astrutbox}{\rule[-5pt]{0pt}{20pt}}

\newcommand\etal{\mbox{\textit{et al.}}}

\newcounter{saveeqn}

\title[Stability boundaries at positive separation ratio]{Stability 
boundaries of roll and square convection in binary fluid mixtures with
positive separation ratio}

\author[B. Huke, M. L\"{u}cke, P. B\"{u}chel, and Ch. Jung]
{B.\ns H\ls U\ls K\ls E, \ns M.\ns L\ls \"{U}\ls C\ls K\ls E,\ns
 P.\ns B\ls \"{U}\ls C\ls H\ls E\ls L,\ns 
\and CH.\ns J\ls U\ls N\ls G}

\affiliation{Institut f\"{u}r Theoretische Physik, Universit\"{a}t des 
Saarlandes,\\ Postfach 151150, D-66041 Saarbr\"{u}cken, Germany}

\pubyear{1999}
\volume{???}
\pagerange{1--100}
\date{?? and in revised form ??}
\begin{document}

\maketitle

\begin{abstract}
Rayleigh--B\'{e}nard convection in horizontal layers of binary 
fluid mixtures heated from below with realistic horizontal boundary conditions
is studied theoretically using multi-mode Galerkin expansions. For positive
separation ratios 
the main difference between the mixtures and pure fluids lies in the 
existence of stable three dimensional patterns near onset in a wide range of 
the parameter space. We evaluated the stationary solutions of
roll, crossroll, and square convection and we determined the location of the
stability boundaries for many parameter combinations thereby obtaining 
the Busse balloon for roll and square patterns.

\end{abstract}

\section{Introduction}
\label{intro}

Convection in binary miscible fluids like ethanol-water, $^3$He 
$-^4$He, or various gas mixtures shows a rich spectrum of 
pattern formation behaviour --- see, e.g., 
(\cite[Platten \& Legros 1984]{PL84}; 
\cite[Cross \& Hohenberg 1993]{CH93};
\cite[L{\"u}cke \etal 1998]{LBBFHJ98})
for a review. The spatiotemporal properties of convection
in mixtures are more complex than those of one-component fluids due to the
influence of Soret sustained concentration gradients. 
The structural dynamics of the concentration distribution in mixtures
results from an interplay between three competing mechanisms: nonlinear 
advection and mixing, weak solutal diffusion, and the Soret effect. The latter
generates and sustains concentration gradients in (linear) response to  
local temperature gradients. Without Soret coupling, i.e., for vanishing
separation ratio $\psi=0$ any concentration fluctuation diffuses away.
For $\psi \neq 0$\,, however, the
externally imposed vertical temperature difference across the fluid layer 
sustains via the Soret
effect concentration variations against the action of advective mixing and 
diffusive dissipation.

The concentration field changes the advective properties of mixtures via 
solutal buoyancy forces that enter into the momentum balance of the fluid. 
Thus, a concentration fluctuation directly influences the flow which in turn 
changes and mixes the concentration. In binary {\it liquids\/}, this 
nonlinear feed back is only weakly damped by diffusive 
homogenisation so that the concentration distribution shows  
anharmonic and boundary layer structures. Furthermore, it is ultimately this 
feed back that causes already right at onset convection patterns that cannot be 
seen there in pure fluids. Examples that occur depending on parameters 
are traveling waves of roll structures, 
standing wave oscillations, and stationary squares. In addition mixtures show
very interesting secondary structures close to onset: spatially
localized traveling wave states, stationary 
crossrolls (CRs), and oscillations between squares and rolls or CRs can be seen
for different parameter combinations. Note that all of the
aforementioned patterns are Soret induced by the concentration field --- they
disappear in the pure fluid limit, $\psi \to 0$,
when switching off the Soret coupling to the concentration field.

In this paper we are concerned with the case of a positive Soret effect, 
$\psi>0$,
which causes the heavier (lighter) component of the mixture to be driven 
towards lower (higher) temperature regions. Therefore, heating a mixture
with $\psi>0$ from
below establishes a stronger density gradient as in a pure fluid. The solutal
contribution to the buoyancy increases the thermal destabilization of the 
fluid layer and convection starts at smaller temperature differences compared
to a pure fluid. One commonly denotes the thermal driving regime with Rayleigh
numbers $R$ below the threshold $R_c^0$ of pure fluids as the Soret regime and
the regime above $R_c^0$ as the Rayleigh region 
(\cite[Moses \& Steinberg 1991]{MS91}). As a crude rule of thumb
one can say that in the Soret region, $R<R_c^0$, square patterns are often 
observed in mixtures, whereas in the Rayleigh region stable rolls are found.

There have been only few theoretical investigations aimed at explaining the 
transition scenario between squares at smaller $R$ and rolls at larger $R$
(\cite[Clune \& Knobloch 1992]{ClK92}; 
\cite[M{\"u}ller \& L{\"u}cke 1988]{ML88}).
Recently we have elucidated this transition for a fixed wavenumber 
(\cite[Jung, Huke, \& L{\"u}cke 1998]{JHL98}). 
In this paper we compare the properties of square, roll, and CR patterns and we 
present a comprehensive linear stability analysis of 
rolls and a more restricted one for squares. 
We elucidate how the stability boundaries of rolls that have been 
determined by Busse and coworkers 
(\cite[Busse 1967]{B67};
\cite[Bolton, Clever, \& Busse 1985]{BBC85};
\cite[Clever \& Busse 1990]{CB90}) for 
pure fluid convection are modified by taking into account the influence of the 
concentration field in mixtures. We present for the first time a full numerical 
investigation of the stability behaviour of rolls and squares and present
the stability balloons of these patterns in the $k-r$--plane for a wide range 
of fluid parameters.    

The paper is organized as follows: In~\S\,\ref{basics} we describe briefly 
the basics of convection in binary fluids and we explain the application of 
the Galerkin method to this particular system. In~\S\,\ref{struct} we describe 
and compare the stationary solutions for rolls, CRs, and squares. 
In~\S\,\ref{stabili} the Galerkin method is used for the linear stability 
analysis of roll and square patterns. 
We conclude in~\S\,\ref{conclude} with a summary of our results.  

\section{Mathematical foundations}
\label{basics}

In this paper we investigate 
convection in horizontal binary fluid layers confined between perfectly
heat conducting, rigid, impermeable plates.
Since the system and its basic equations are well known 
(\cite[Landau \& Lifshitz 1966]{LL66}; 
\cite[Platten \& Legros 1984]{PL84}),
we summarize in~\S\,\ref{sec21} only the necessary formulas for our 
investigation. Then we present relevant details related to the application of 
the Galerkin expansion technique to this system.

\subsection{System and basic equations}
\label{sec21}

We consider a horizontal layer of a binary fluid mixture of thickness $d$ in
a homogeneous gravitational field, ${\bf g} = - g \, {\bf e}_z$.
A vertical temperature gradient is imposed by fixing the temperature
\begin{equation}
T = T_0 \pm \frac{\Delta T}{2} 
\,\,\, \mbox{at} \,\,\, z = \mp \frac{d}{2} \, ,
\end{equation}
e.g., via highly conducting plates in experiments. Here we consider the 
plates to be infinitely extended, rigid, and impermeable.

Convection is described in terms of the fields of velocity ${\bf u} =(u,v,w)$, 
temperature $T$, mass concentration $C$ of the lighter component, total mass 
density $\rho$, and pressure. In the balance equations connecting 
these fields we scale lengths and positions by $d$, time by the 
vertical thermal diffusion time $d^2 / \kappa$, temperature by 
$\nu \kappa / \alpha g d^3$, concentration by $\nu \kappa / \beta g d^3$, 
and pressure by $\rho_0 \kappa^2/d^2$. Here $\rho_0$ is 
the mean density, $\kappa$ the thermal diffusivity, $\nu$ the kinematic 
viscosity, and $\alpha$ and $\beta$ are thermal and solutal expansion
coefficients, respectively.
Using the Oberbeck-Boussinesq approximation 
the balance equations read 
(\cite[Platten \& Legros 1984]{PL84}; 
\cite[Hort, Linz, \& L{\"u}cke 1992]{HLL92})
\begin{subeqnarray}
{\bf \bnabla\cdot u} & = & 0 \slabel{ground21}\\
(\partial_t + {\bf u\cdot\bnabla\,})\,{\bf u} & = & - {\bf\bnabla}\,p  +  
\sigma\,\left[\,\left(\theta + c\,\right)\,{\rm\bf e}_z + 
\bnabla^2\,{\bf u}\,\right]\slabel{ground22}\\
(\partial_t + {\bf u\cdot\bnabla\,})\,\theta & = & 
R\,w  + \bnabla^2\,\theta\slabel{ground23}\\
(\partial_t + {\bf u\cdot\bnabla\,})\,c & = & R\,\psi\,w  +  
L\,\left(\,\bnabla^2\,c - \psi\,\bnabla^2\,\theta\,\right)\slabel{ground24} \;\; .
\end{subeqnarray}
Here $\theta, c$, and $p$ are the reduced deviations of temperature, 
concentration, and pressure, respectively, from the conductive profiles.                              

The Lewis number $L$ is the ratio of the
concentration diffusivity $D$ to the thermal diffusivity $\kappa$, therefore
measuring the velocity of concentration diffusion. The Prandtl number 
$\sigma$ is the ratio of the momentum diffusivity $\nu$ and $\kappa$:
\begin{equation}
L = \frac{D}{\kappa}\,\,;\,\, \sigma = \frac{\nu}{\kappa} 
\label{GdefLsig} \,.
\end{equation}
The Rayleigh number $R$ measures the thermal driving and the separation ratio
$\psi$ measures the strength of the Soret coupling between temperature and 
concentration fields
\begin{equation}
R =\frac{ \alpha g d^3 \Delta T}{\nu \kappa}\,\,;
\,\psi = - \frac{\beta}{\alpha} \frac{k_T}{T_0}\, .
\label{GdefR} 
\end{equation}
Here $T_0$ is the mean 
temperature and $k_T$ is the thermal diffusion ratio 
(\cite[Landau \& Lifshitz 1966]{LL66}). The driving forces entering into 
the momentum balance equation (\ref{ground22}) are pressure 
gradients and the 
buoyancy caused by the temperature and concentration dependence of the 
density. 

The off-diagonal term 
$-L\,\psi\,\bnabla^2\,\theta\,$ and the term $R\,\psi\,w$
in the concentration balance equation (\ref{ground24})
describe the action of the Soret effect, i.e., the generation of concentration 
currents and concentration gradients by temperature 
variations. A Soret coupling $\psi > 0$ implies a positive Soret effect. 
In this case the
lighter component of the mixture is driven into the direction of higher 
temperature thus increasing the density variations. 

The Dufour effect, i.e., the driving 
of temperature currents by concentration variations is of interest only in 
gas mixtures (\cite[Hort \etal 1992]{HLL92}). But even there 
it is often small (\cite[Liu \& Ahlers 1997]{LA97}). 

\subsection{Galerkin method}
\label{sec22}

To describe three--dimensional patterns with wavenumbers $k_x$ and $k_y$ each 
field $X$ is expanded as
\begin{equation}
X(x,y,z;t) = \sum_{lmn} X_{lmn}(t) \mathrm{e}^{\mathrm{i} l k_x x} 
\mathrm{e}^{ \mathrm{i} m k_y y } f_n(z) \;\; .
\label{expansion}
\end{equation}
Here $l$ and $m$ are integers and the $f_n$ form a complete system of 
functions that fits the specific boundary condition for the field $X$ at the
plates. 
To find suitable sets of functions $f_n$ we introduce some new fields.
First, two scalar fields $\Phi$ and $\Psi$ are defined via
\begin{equation}
{\bf u} = {\bf \bnabla} \times {\bf \bnabla} \times \Phi {\bf e}_z
+ {\bf \bnabla} \times \Psi {\bf e}_z \;\; . \label{22phipsidef}
\end{equation}
The structures we want to discuss do not show a horizontal mean flow for 
mirror symmetry reasons. Then, (\ref{22phipsidef})
is the most general expression that fulfills the incompressibility
condition (\ref{ground21}) (\cite[Clever \& Busse 1989]{CB89}).
The analysis of mean flow effects in the perturbations in discussed in 
section~\ref{sec24}.

Second, instead of $c$ we use the field 
\begin{equation}
\zeta = c - \psi \theta
\end{equation}
that allows in a more convenient way to guarantee the impermeability of 
the horizontal boundaries: The diffusive part of the concentration current, 
driven by concentration gradients as well as by temperature
gradients is given by $- L {\bf \bnabla} (c - \psi \theta)$. 
At the impermeable plates the vertical component of this current vanishes 
which requires
\begin{equation}
0 = \partial_z \left(c - \psi \theta \right) = 
    \partial_z \zeta  \hspace{1cm} 
\mbox{at } z = \pm 1/2 \;\; .
\end{equation}
The advective concentration current vanishes at the plates because there 
${\bf u} = 0$.
The balance equation for $\zeta$ is obtained by combining (\ref{ground23}) and
(\ref{ground24}).

The boundary conditions for the fields $\Phi, \Psi, \theta$, and 
$\zeta$ read
\begin{equation}
\Phi = \partial_z \Phi = \Psi = \theta = \partial_z \zeta = 0 
\hspace{2cm} \mbox{at } z = \pm 1/2 \;\; .
\end{equation}
To expand the fields 
$\Psi$, $\theta, \zeta$, and $\Phi$ vertically we used different 
orthonormal sets $f_n(z)$ as follows
\begin{subeqnarray}
\Psi \mbox{\, and \,} \theta \,:\,f_n(z) &=& \, 
\left\{ \begin{array}{ll}
\sqrt{2} \cos(n \upi z)  & \mbox{\qquad $n$ odd} \\ 
\sqrt{2} \sin(n \upi z)  & \mbox{\qquad $n$ even} 
\end{array} \right.
\slabel{csdefinition} \\
\zeta \,:\, f_n(z)  &=& \left\{ \begin{array}{ll}
1 & \mbox{\qquad  $n = 0$}\\
\sqrt{2} \, \sin(n \upi z) & \mbox{\qquad $n$ odd} \\
\sqrt{2} \, \cos(n \upi z) & \mbox{\qquad $n \neq 0$ even}
\end{array}
\right.
\slabel{scdefinition} \\
\Phi \,:\, f_n(z) &=& \left\{ \begin{array}{ll} 
C_{\frac{n+1}{2}}(z) & \mbox{\qquad $n$ odd} \\ 
S_{\frac{n}{2}}(z) & \mbox{\qquad $n$ even} 
\end{array} \right. \;\; .
\slabel{CSdefinition}
\end{subeqnarray}
Here $C_n$ and $S_n$ are Chandrasekhar functions 
(\cite[Chandrasekhar 1981]{C81}).

The balance equations for the new fields are
\begin{subeqnarray}
\partial_t \Delta_2 \Psi & = & 
\sigma {\bf \nabla}^2 \Delta_2 \Psi + 
\left\{{\bf \bnabla} \! \times \left[ \left({\bf u} \cdot \bnabla \right) 
{\bf u} \right] \right\}_z  \slabel{ground71}\\
\partial_t {\bf \nabla}^2 \Delta_2 \Phi & = & 
\sigma \left\{ {\bf \nabla}^4 \Delta_2 \Phi -  
\Delta_2 \left[ \left( 1 + \psi \right) \theta + \zeta \right] \right\} 
- \left\{{\bf \bnabla} \! \times {\bf \bnabla} \! \times
\left[ \left( {\bf u} \cdot \bnabla \right) {\bf u} \right] \right\}_z 
\slabel{ground72} \\
(\partial_t + {\bf u\cdot\bnabla\,})\,\theta & = & 
-R\, \Delta_2 \Phi  + \nabla^2\,\theta \slabel{ground73} \\
(\partial_t + {\bf u\cdot\bnabla\,})\,\zeta & = & 
L \nabla^2 \zeta - \psi \nabla^2 \theta \slabel{ground74}
\;\; .
\end{subeqnarray}
Here $\Delta_2 = \partial_x^2 + \partial_y^2$.

By inserting the ansatz (\ref{expansion}) for each field into the balance 
equations and projecting them 
onto the basic functions one gets a nonlinear algebraic system of equations of 
the form
\begin{equation}
A_{\kappa \mu} \partial_t X_\mu = B_{\kappa \mu} X_\mu + 
C_{\kappa \mu \nu} X_\mu X_\nu \;\; . \label{basicslinsyseq}
\end{equation} 
For simplicity amplitudes are labelled here by a single Greek 
index and the summation convention is implied in (\ref{basicslinsyseq}) with
$A_{\kappa \mu}, B_{\kappa \mu}$, and $C_{\kappa \mu \nu}$ being constant
 coefficients.
 
The number of modes has to be truncated to get a 
finite number of equations as discussed later on. For stationary 
convection structures the 
left hand side of (\ref{basicslinsyseq}) vanishes and the solution can be 
found using a multidimensional Newton method.

\subsection{Symmetries}
\label{sec23}

Symmetries of convective structures impose conditions on the fields and/or 
imply relations between different
modes of the fields thereby restricting the number of independent modes that 
are necessary to describe the patterns. For example, to describe
two--dimensional roll 
patterns with $k_x = k$, $k_y = 0$ that do not depend on $y$, all amplitudes
with $m \neq 0$  are set to zero in (\ref{expansion}). On the other hand,
square patterns are 
characterized by $k_x = k_y = k$ and $X_{lmn}=\pm X_{mln}$. But we also 
investigate three--dimensional CR
patterns with $k_x = k_y = k$ for which, however, $X_{lmn} \neq \pm X_{mln}$.

\subsubsection{Stationary rolls}

To describe these two--dimensional structures one does not need the 
$\Psi$--field.
Furthermore, rolls are even in $x$ with an appropriate choice of the plane
$x=0$. As a consequence of this
mirror symmetry one has  $X_{l0n} = X_{-l0n}$ so that the lateral 
functions $\mathrm{e}^{\pm \mathrm{i} k l x}$ can be replaced by 
$\cos (k l x)$. In addition the roll pattern is antisymmetric under reflection 
at the plane $z=0$ combined with a translation by half a wavelength in 
$x$-direction. This mirror glide symmetry enforces 
half of the amplitudes to be zero, e.~g.\ all amplitudes 
$\Phi_{l0n}$ where $l+n$ is an odd number.

It is no accident that stationary roll patterns have these symmetries. 
That they are
fulfilled at onset can be shown via a linear stability analysis of the 
conductive state (\cite[Hollinger \& L{\"u}cke 1995]{HL95}). The subset of 
modes that obey these symmetries is closed in the sense that these modes do 
not drive others via nonlinear coupling. Thus, the observed roll solution 
remains symmetric as long as no symmetry breaking bifurcation occurs on the 
stationary roll branch with symmetry breaking modes becoming linearly unstable.
Such instabilities are covered by our stability analysis.
\cite[Moore, Weiss, \& Willkins (1991)]{MWW91} have discussed these symmetry 
breaking
perturbations for free--slip and permeable boundary conditions. 

\subsubsection{Stationary squares and crossrolls with $k_x = k_y$}

They have the same symmetry plane at $x=0$ as rolls and an 
additional mirror plane at $y=0$. Furthermore, the squares and CRs
show also a mirror glide symmetry. Here, however, the symmetry transformation
consists of a reflection at
the plane $z=0$ combined with a translation by half a wavelength in 
$x$- {\em and} $y$-direction. To describe these three--dimensional patterns 
the $\Psi$--field
cannot be neglected. We also mention that in contrast to the other fields
 $\Psi$ is odd in $x$ and $y$ and has 
positive parity under the mirror glide operation thereby reflecting the 
symmetries of the velocity field. 

For square patterns that are invariant under rotation by $90^{\circ}$
in which the $x$- and $y$-directions are indistinguishable
a further reduction of the number of mode occurs: amplitudes like 
$\Phi_{lmn}$ and $\Phi_{mln}$ are the same. This is also true for $\theta$
and $\zeta$. Again $\Psi$ is different. Here $\Psi_{lmn}=-\Psi_{mln}$.

\subsection{Stability analysis}
\label{sec24}

To make a full stability analysis one has to check the stability of the 
patterns against perturbations with arbitrary wavevector 
$d {\bf e}_x + b {\bf e}_y $. To do so one has
to introduce a Floquet term writing out the perturbation as
\begin{equation}
\delta X(x,y,z;t) = \mathrm{e}^{\mathrm{i} d x + 
\mathrm{i} b y} \mathrm{e}^{s t}
\sum_{lmn} \delta X_{lmn}  \mathrm{e}^{\mathrm{i} l k_x x + 
\mathrm{i} m k_y y} f_n(z).
\end{equation}
Such a perturbation is added to the known solution the stability of which is to 
be tested and inserted into the 
balance equations. After linearising and projecting one gets a linear eigenvalue
problem of the form
\begin{equation}
s A_{\kappa \mu} \delta X_\mu = B_{\kappa \mu} \delta X_\mu \;\; . 
\end{equation}
with constant coefficients $A_{\kappa \mu}$ and $B_{\kappa \mu}$. 
The aforementioned solution, i.e., the convective structure described by it 
is stable if 
every eigenvalue $s$  has a negative real part for every $d$ and $b$.

The symmetry of the convective pattern discussed above can under some 
circumstances be 
used to get separated classes of possible eigenvectors representing the 
perturbations. That means the eigenvalue problem can be reduced 
to finding the eigenvalues of two matrices of about half of the size. Because 
evaluating the eigenvalues of a matrix is a $O(N^3)$-process this always implies 
a reduction of the computation time.

\subsubsection{Stationary rolls}
\label{sec241}

To perform the stability analysis of rolls one determines the growth behaviour
of perturbations of the form
\begin{equation}
\delta X(x,y,z;t) =  \mathrm{e}^{\mathrm{i} d x + \mathrm{i} b y} 
\mathrm{e}^{s t} \sum_{ln} \delta X_{l0n}  \mathrm{e}^{\mathrm{i} l k x} f_n(z).
\label{perturbations}
\end{equation}

Contrary to the roll solutions, their perturbations may contain a mean flow
component that is discarded in (\ref{22phipsidef}). However, 
our perturbation ansatz (\ref{perturbations}) contains modes like 
$\delta \Phi_{001} \mathrm{e}^{\mathrm{i} d x + \mathrm{i} b y} C_1(z)$
describing a mean flow in the limit $b,d \rightarrow 0$, and the equation
of motion for $\delta \Phi_{001}$ reduces to two independent equations
for the mean flow as they are used e.~g. in
(\cite[Clever \& Busse 1991]{CB91}).
If there is a mean 
flow in the perturbations, then modes like the $\delta \Phi_{001}$--mode 
diverge when $b,d$ goes to zero such that the long wavelength component of 
the velocity remains finite.
However, in the parameter range explored here, the perturbations limiting the
stability balloon of rolls have finite $b$ or $d$ and thus no mean flow.

Because of the periodicity of the patterns in $x$ and its mirror symmetry 
it suffices to consider $d \in \left[ 0, k/2 \right]$. In 
$y$-direction, however, all perturbation wavenumbers, say, $b \ge 0$ 
have to be investigated.

The linear system of 
equations (\ref{perturbations}) always separates into two subsystems of
perturbations $\delta X_\mu$ that belong to modes with amplitudes $X_\mu$ that 
are antisymmetric ($G$--perturbations) or symmetric 
($\overline{G}$--perturbations) under the mirror glide operation 
$(x, y, z) \rightarrow (x + \lambda/2, y + \lambda/2, -z)$.
E.~g.\ all 
perturbations with amplitudes $\delta \Phi_{l0n}$ with even $l+n$ are 
$G$--perturbations, and the perturbations with odd $l+n$ are 
$\overline{G}$--perturbations.

In general the perturbations (\ref{perturbations}) do not have a well 
defined symmetry under the 
mirror glide operation. This is only the case if $d = n k$. If $n$ is even 
(odd) then the perturbations have the same (opposite) parity as the modes they
belong to. Every $G$--perturbation can be written out as a 
$\overline{G}$--perturbation (and vice versa) by choosing a new $d' = d - k$.
Thus the distinction between $G$-- and $\overline{G}$--perturbations is well defined 
only for fixed $d$.

It is possible to restrict the stability analysis to {\em one} set of
perturbations by extending the investigated $d$-interval to $[0,k]$. 
Consider for example a $\overline{G}$--perturbation with 
$d \in \left[0, k/2 \right]$. It can as well be written as 
a $G$--perturbation  with a new $d' = d - k$ or equivalently 
$d' = k - d$. Thus one finds all $\overline{G}$--perturbations with 
$d \in \left[0, k/2 \right]$ again as $G$--perturbations with 
$d' \in \left[k/2, k \right]$. We will therefore restrict ourselves to 
the set of perturbations that has negative parity under the mirror glide
operation at $d=0$, i.~e.,\ we investigate the $G$--perturbations in the whole
interval $\left[0, k\right]$.

In special cases the system of equations can be separated even further. For
$d=0$ the perturbations can be divided into those that are symmetric and those
that are antisymmetric under the operation $x \rightarrow -x$. Furthermore, 
if $b=0$ then the perturbations contain either no or only
$\delta \Psi$-amplitudes.

\subsubsection{Stationary squares and CRs with $k_x = k_y$}

Because the amount of computational power needed to make a full stability 
analysis of these three--dimensional structures is too large, we will discuss 
perturbations only for periodic boundary conditions, i.~e.,\ 
$d=b=0$. Here again a separation of perturbations is possible into those 
that change sign or not under the mirror glide operation.

Furthermore, the stability problem is invariant under $x \rightarrow -x$ and 
$y \rightarrow -y$. Thus one can distinguish between perturbations that are
even in $x$ and $y$, odd in $x$ and $y$, or even in $x$ and odd in $y$ (or
equivalently odd in $x$ and even in $y$). If the perturbations have the same
symmetry in both directions one can in the case of squares finally make use of 
a last symmetry property and separate between perturbations that are symmetric
or antisymmetric under the exchange of the $x$- and $y$-direction. This is not
possible for CR patterns because of their lack of the 
$x \leftrightarrow y$ symmetry. 

As we will see in section~\ref{sec43}, the destabilising perturbations of
squares fall into the subclass that is even in $x$ and $y$, and therefore do not
drive a mean flow. 

\section{Properties of the patterns}
\label{struct}

Squares, CRs, and rolls are realized as stable convection 
structures somewhere in parameter space. Furthermore, if $L$ is sufficiently
small these
three patterns appear for fixed $L$, $\sigma$, $\psi$, and $k$ as global 
attractors 
at different Rayleigh numbers (however, CRs coexist bistably with
oscillations in a small $R$--interval (\cite[Jung \etal 1998]{JHL98})). So it 
is easy to enforce their stable 
experimental realization successively by increasing the Rayleigh number 
beyond the onset of convection. 

To understand the 
behaviour of the patterns it is useful to consider the driving region near 
onset and the Rayleigh region separately. 
In the pure fluid the critical point is at a Rayleigh number 
$R_c^0 =R_c(\psi=0)=1707.762$ and a wavenumber
$k_c^0 =k_c(\psi=0) = 3.117$ (\cite[Chandrasekhar 1981]{C81}). In binary 
mixtures 
with positive separation ratios that we are dealing with, the critical Rayleigh
number is smaller, $R_c(\psi>0) < R_c^0$, since the solutal contribution
to the quiescent state's buoyancy force enhances the latter. Thus a smaller
thermal driving, i.e., a smaller  Rayleigh number suffices to reach the
critical buoyancy force size for onset of convection.
The critical wavenumber is also somewhat lower: $k_c(\psi>0) < k_c^0$ 
(\cite[Knobloch \& Moore 1988]{KM88}). 

When presenting our results we shall use the reduced  
Rayleigh number 
\begin{equation}
r = \frac{R}{R_c^0}
\end{equation}
and the reduced distance 
\begin{equation}
\epsilon = \frac{R}{R_c} -1 = \frac{r}{r_c}-1.
\end{equation}
from threshold $r_c=\frac{R_c}{R_c^0}$

\subsection{Small-amplitude convection --- amplitude equation}
\label{sec31}

Very close to onset the wavenumber dependent bifurcation behaviour of rolls 
and squares can in pure fluids 
as well as in  binary mixtures be described most simply by
two coupled cubic amplitude equations of the form
\begin{subeqnarray}
\tau_0\partial_t A & = & \epsilon A +
\xi_0^2\left(\partial_x - \frac{i}{2 k_c}\partial_y^2 \right)^2 A - |A|^2 A
-f |B|^2 A \slabel{ampligleich1} \\
\tau_0\partial_t B & = & \epsilon B + 
\xi_0^2\left(\partial_y - \frac{i}{2 k_c}\partial_x^2 \right)^2 B - |B|^2 B
-f |A|^2 B  \slabel{ampligleich2} \;\; .
\label{ampligleich}
\end{subeqnarray}
Clune and Knobloch used such equations without the spatial derivative term
(\cite[Clune \& Knobloch 1991]{ClK91}). For a review of such amplitude equations
and how they are related to the basic equations see, e.~g., 
(\cite[Cross \& Hohenberg 1993]{CH93}). 

The amplitudes $A$ and $B$ of the eigenfunctions of the linearized
hydrodynamic field equations correspond quite well to the leading 
amplitudes 
\begin{subeqnarray}
w_{101} = k^2 \Phi_{101} &=& k^2 \int 
\,\Phi(x,y,z)\,\mathrm{e}^{\mathrm{i}kx}\, 
\, C_1(z) \, \mathrm{d}x\mathrm{d}y\mathrm{d}z \\
w_{011} = k^2 \Phi_{011} &=& k^2 \int 
\,\Phi(x,y,z)\,\mathrm{e}^{\mathrm{i}ky}\, 
\, C_1(z) \, \mathrm{d}x\mathrm{d}y\mathrm{d}z \;\;.  
\end{subeqnarray}
of the vertical velocity field or the corresponding amplitudes of the 
temperature field. 

The two roll solutions of (\ref{ampligleich}) are
$A=|A_R|\mathrm{e}^{\mathrm{i}(k-k_c)x}, B=0$ for rolls with wavevector 
${\bf k} = k {\bf e}_x$ and 
$A=0, B=|B_R|\mathrm{e}^{\mathrm{i}(k-k_c)y}$ for rolls with wavevector 
${\bf k} = k {\bf e}_y$.
Squares are described by the solution 
$A=|A_S|\mathrm{e}^{\mathrm{i}(k-k_c)x}, 
B=|B_S|\mathrm{e}^{\mathrm{i}(k-k_c)y}$ with $A_S=B_S$.
Crossrolls, i.~e.\ solutions with finite $|A| \neq |B|$ do not exist in 
(\ref{ampligleich}).

If $f > -1$, the square solution exists besides the roll solution
for $\epsilon$ above the 
neutral stability curve
\begin{equation}
\epsilon_{stab}(k) = \xi_0^2 (k - k_c)^2  \;\; .
\end{equation}
We will always consider roll patterns with $B_R = 0$ and $A_R\neq 0$.
For the roll solutions one has 
\begin{equation}
|A_R|^2 = \epsilon - \epsilon_{stab}(k) \;\; ,
\end{equation}
and for squares one finds
\begin{equation}
|A_S|^2 = |B_S|^2 = \frac{\epsilon - \epsilon_{stab}(k)}{1+f} \;\; 
\end{equation}
so that
\begin{equation}
|A_R|^2 = (1+f)|A_S|^2\;\; .
\label{amplirelation}
\end{equation} 

A stability analysis of the roll and square solutions shows that
squares (rolls) are stable (unstable) whenever $|A_R|^2 < 2|A_S|^2$, i.e., for 
$-1 < f < 1$ and they are unstable (stable) whenever $|A_R|^2 > 2|A_S|^2$.

\subsection{Full Galerkin expansion}
\label{sec32}

Both square and roll solutions exist for every $r$ above onset $r_c(\psi,L)$
and the mode intensity $|w_{101}|^2$ for rolls is always greater than
for squares. Hence the parameter $f$ in the amplitude 
equation has to be greater than $0$ according to (\ref{amplirelation}).
In figure \ref{flow_intensity} we present  
a plot of the leading contribution $|w_{101}|^2+|w_{011}|^2$
to the vertical flow intensity $w^2$ versus $r$ as obtained from the 
full Galerkin expansion for several parameters.

By comparing our results from the full Galerkin expansion with the 
amplitude equation approximation we verified 
that the latter works well near onset 
$r_c(\psi)$ which in figure \ref{flow_intensity} lies significantly 
below 1 -- outside the plot range of figure \ref{flow_intensity}. 
However, at larger $r$ when approaching the Rayleigh regime the 
amplitude approximation loses its validity:
$|A|^2 + |B|^2$ continues to grow linearly with $r$ 
with its initial slope at onset whereas
$|w_{101}|^2 + |w_{011}|^2$ strongly curves upwards in the 
Rayleigh regime. Furthermore, while
$|w^{R}_{101}|^2 < |w^{S}_{101}|^2 + 
|w^{S}_{011}|^2 = 2 |w^{S}_{101}|^2$
holds close to onset --- as required within the amplitude equation 
approximation for squares to be stable --- the full solutions at larger 
$r$ are such that
$|w^{R}_{101}|^2 > |w^{S}_{101}|^2 + |w^{S}_{011}|^2$ 
without, however, rolls becoming stable, what happens at even higher $r$. 

In the Rayleigh region the 
amplitudes become comparable with those of the pure fluid. This is because the 
concentration field gets more and more advectively mixed and equilibrated 
and loses 
therefore its influence on the convection. Pure fluid 
convection, $\psi=0$, can be described close to onset $r_c(\psi=0)=1$ 
by a cubic amplitude equation. However, the initial slope of $w^2$ versus $r$
is much greater than for the binary mixtures shown in figure \ref{flow_intensity} 
with $\psi>0$. The transition between Soret and 
Rayleigh region is especially sharp at small $L$ -- c.f. the bifurcation
diagram for $L=0.01$ in figure \ref{flow_intensity}(b). We will observe such a 
behaviour again when we consider the stability of the patterns. 

In the $r$-range investigated in this paper the field amplitudes 
$\Phi_{101}$ and $\Phi_{011}$ that are kept in the amplitude equation 
approximation are also the leading ones in the full Galerkin expansion. 
To describe only the fixed point solutions it would
indeed be sufficient to approximate the velocity field of rolls (squares and
CRs) by one mode (two modes) only. But this restriction does not suffice
when the linear stability is investigated. 

Although more than two amplitudes are needed, a good representation is easier to 
achieve for the velocity field than for the $\zeta$-- and
$\theta$--fields. For them very many modes are needed  if $L$ is small and 
$r$ is large. This is a region of the parameter space where the concentration 
field shows narrow boundary layer behaviour which has to be resolved properly. 
In addition a consistent description of the temperature field requires then
also --- despite the fact that it is rather smooth --- high $\theta$--modes
as discussed by (\cite[Hollinger \& L{\"u}cke 1998]{HL98};
\cite[Hollinger 1996]{H96}). 

We followed
(\cite[Clever \& Busse 1989]{CB89}) when defining our truncation 
prescription for the Galerkin expansion. We defined a maximal mode index $N$ and 
neglected all modes $X_{lmn}$ with $|l|+|m|+n > N$. We took the smooth 
behaviour of the velocity field into account by defining two different 
indices:  $N_1$ for the $\Phi$-- and  $\Psi$--fields and 
$N_2 = 2 N_1$ for the $\theta$-- and $\zeta$--fields. For the 
most anharmonic roll structures at $r \approx 1.5$, $L < 0.01$, and 
$\psi = 0.15$ that we have investigated expansions up to $N_2=40$ were needed. 
This is much more
than for pure fluids, where truncations with $N \leq 8$ are sufficient to
describe the stability behaviour quantitatively even at large $r$.
Since the structure of squares is somewhat smoother than the one of rolls
(cf.~\S\,\ref{sec34}) and
since they exist stably only at small $r$ in or 
near the Soret region, we needed only $N_2 \leq 20$ for squares. 
Also the CR structures could be described well with such a 
truncation near the first 
bifurcation point at small $r$, where they behave square--like. But in order to
resolve the CR structures also close to 
the second bifurcation point at larger $r$ where the CR solution 
merges into the roll solution (cf.~\S\,\ref{sec33})
more modes would have been necessary.

\subsection{Bifurcation behaviour of rolls, squares, and crossrolls}
\label{sec33}

Figure \ref{bifdiagram} shows a typical bifurcation diagram for a parameter 
combination where the three stationary patterns can be found. It 
also contains information on the stability of these patterns.
Crossrolls exist only in a finite $r$--interval. The CR solution 
branches  emerge out of the square branch slightly above $r = 1$.
At $r\simeq 1.36$ the CR solution disappears when, e.g., the
amplitude $w_{011}^{CR}$ (downwards pointing triangles in 
figure \ref{bifdiagram}) becomes zero and the CR branch for 
$w_{101}^{CR}$ (upwards pointing triangles) ends on the roll solution 
branch $w_{101}^{R}$.

On the other hand, roll as well as square solutions exist for all 
$r \ge r_c$ of figure \ref{bifdiagram}. Initially at onset the latter are 
stable and the former are unstable. For the parameters of figure \ref{bifdiagram}
squares lose their stability in a Hopf bifurcation at $r\simeq 1.11$ to 
oscillations
which eventually undergo with increasing $r$ a subharmonic bifurcation
cascade that is terminated when the CR states have become 
sufficiently attractive to quench the oscillations. For other parameter
combinations, in particular for larger $L$ there are no oscillations and 
the squares transfer their stability directly to CRs
(\cite[Jung \etal 1998]{JHL98}). 

\subsection{Structural properties of roll and square fields}
\label{sec34}

In figure \ref{shadowgraph} we show the concentration distribution 
of square convection for two parameter combinations that are representative
for liquid and gas mixtures. This plot and the concentration 
field structure of 
rolls and squares in a vertical cross section shows a characteristic boundary
layer and plume behaviour at small $L$. Such structures occur when 
advective mixing is large compared to diffusion in the bulk of the fluid.
Consequently the boundary layers and plumes are more pronounced in rolls than 
in squares since $w_R^2 > w_S^2$ as discussed in~\S\,\ref{sec32}.
Thus squares with their broader boundary layers are much smoother structures 
than rolls at the same parameters.

The practically harmonic velocity and temperature 
fields are not shown. For squares they resemble the fields of a linear 
superposition of two perpendicular sets of rolls. 

The Nusselt number $N$ is roughly the same for rolls and squares. Close 
to onset 
\begin{equation}
N-1 \propto |w_{101}|^2 + |w_{011}|^2 \;\; ,
\end{equation}
and the stable structure has the higher Nusselt number there, i.e., 
$N_R<N_S$ in accordance with the inequalities of~\S\,\ref{sec31} predicted 
by the amplitude equation.
Further away from onset, however, one has $N_S<N_R$ thus reflecting the 
magnitude relations of $w^2$ discussed in~\S\,\ref{sec32}.

In figure \ref{mixparam} we show the mixing parameter
\begin{equation}
\label{M}
M = \frac{\sqrt{ \left< \delta C^2 \right> }}
{\sqrt{ \left< \delta C_{cond}^2 \right> }} \;\; .
\end{equation} 
It is defined by the mean square of the deviation, 
$\delta C = C - C_0$, of the concentration from the spatial mean,
$C_0=\left< C \right>$, reduced by the concentration variance in the
quiescent conductive state. Note that $M$ is nearly the same
for the very different concentration fields of rolls and squares, if $L$ is not
too large (figure \ref{mixparam}).

\section{Linear stability analysis of rolls and squares}
\label{stabili}

\subsection{Instability mechanisms of rolls}
\label{sec41}

The stability boundaries of roll patterns in pure fluids are known since the
pioneering work of (\cite[Busse 1978]{B78}).
At small Rayleigh numbers there exist five different instability
mechanisms giving rise to five different stability boundaries 
that limit the region of stable rolls in the $(R,k,\sigma)$--parameter space. 
At small Prandtl numbers the Eckhaus (EC), the skewed varicose (SV), and the 
oscillatory mechanism (OS) are the important instabilities 
(\cite[Clever \& Busse 1990]{CB90}). 
At higher Prandtl numbers the zigzag (ZZ) and the CR mechanisms dominate 
(\cite[Busse 1967]{B67}). Properties and symmetries of these perturbations 
are discussed in (\cite[Bolton \etal 1985]{BBC85}). All 
these five instabilities of roll patterns can also be found in binary mixtures. 

In pure fluids and binary mixtures there exist always perturbations 
of the form (\ref{perturbations}) with 
$b=d=0$ against which a roll pattern is only marginally stable. Such a 
perturbation 
has no $\Psi$--component and is odd in $x$. It reflects just an infinitesimal 
shift of the whole pattern in the $x$--direction. Therefore  such a 
perturbation has an eigenvalue $s=0$. For later discussion we point out here 
that this particular eigenvalue is connected to nearly all instabilities:
the perturbations causing them and the associated eigenvalues evolve
smoothly into the lateral shift when one moves in the $d-b$ plane  
from the $d-b$ coordinates that locate the instability to the origin $d=b=0$.

In the remainder of ~\S\,\ref{sec41} we will briefly characterize the 
properties of
the aforementioned five perturbations before presenting our results of
the stability analyses for rolls in ~\S\,\ref{sec42} and for squares in
~\S\,\ref{sec43}. We begin with 
the three types (EC, ZZ, and CR) that touch the critical point 
$\epsilon = 0$, $k = k_c$ and that 
exist also within the 
two coupled amplitude equations (\ref{ampligleich1} - \ref{ampligleich2}).

\subsubsection{Eckhaus instability}
\label{sec411}

Perturbations of the Eckhaus type are most critical at $b = 0$ 
and have no $\Psi$--component. Thus they can be described as 
purely two--dimensional.
The EC instability tends to establish a new set of rolls with a better 
wavenumber in the direction of the wave vector of the unstable roll pattern.
Within the amplitude 
equations EC perturbations can be represented as variations of the
$A$--amplitude that depend only on $x$. Instability occurs here for 
$\epsilon$ and $k$ such that 
$\epsilon_{stab}(k) < \epsilon < \epsilon_{EC}(k)= 3 \, \epsilon_{stab}(k)$.
At $d \rightarrow 0$ the EC perturbations reduce both in the amplitude equation
approximation and in the full hydrodynamic equations to the lateral shift.
For symmetry reasons $s \sim d^2$ near $d = 0$. In the case of EC 
instability (stability) there is a minimum (maximum) of $s$ at 
$d =0$. It is therefore sufficient to investigate the pattern at a 
single point on the $d$--axis near $d = 0$ numerically in order to determine
the stability behaviour
against EC perturbations. But to find the most critical value of $d$ an
evaluation and interpolation of $s$ along the $d$--axis is necessary. 

\subsubsection{Zigzag instability}
\label{sec412}

Zigzag perturbations have $d=0$ and fall into the subclass of 
perturbations that are odd in $x$. In the amplitude equations they show up
as $y$--dependent perturbations in the $A$--amplitude.
They cause the growth of a new set of rolls that has always a 
greater wavenumber than the original set. Consequently they confine the region 
of stable rolls on the small--$k$ side. The amplitude 
equations predict a ZZ instability for all $k < k_c$. Like EC
perturbations the ZZ instability reduces to the lateral shift when $b
\rightarrow 0$ and the question of stability can be answered at a single point
near $d=b=0$.

\subsubsection{Crossroll instability}
\label{sec413}

It occurs when roll--like perturbations perpendicular 
to the existing pattern can grow. 
In the amplitude equations (\ref{ampligleich1} - \ref{ampligleich2}) they 
are described as perturbations in the amplitude $B$ when $A$ describes the
stationary roll pattern. Rolls are CR-unstable for 
$\epsilon_{stab} (k) < \epsilon < \epsilon_{CR} (k) 
= f/(f-1) \,\, \epsilon_{stab} (k)$
when $f > 1$. In this case $\epsilon_{CR} (k)$ can be above or below
the Eckhaus boundary 
$ \epsilon_{EC} (k) = 3\, \epsilon_{stab} (k)$ depending on whether
$f < 3/2$ or not.
However, if $f < 1$ 
(which is the case when squares are stable --- c. f. ~\S\,\ref{sec31})
then rolls are $CR$--unstable for all
$\epsilon > \epsilon_{stab} (k)$ within the amplitude
equations approximation (\ref{ampligleich1} - \ref{ampligleich2}).
Note that the case $f < 1$ does not occur in pure fluids but 
it can occur in binary mixtures when squares are stable at onset. 

In the multi--mode Galerkin expansion the
leading mode in the velocity field of the CR perturbation has the form
\begin{equation}
\delta \Phi_{001} e^{iby} C_1(z) \;\; , \label{CRmode}
\end{equation}
where near the critical point $b \approx k_c$. 
This is a mode of the $\overline{G}$ class of perturbations that are
symmetric under the mirror glide operation as discussed in 
~\S\,\ref{sec241}. Being a member of the $\overline{G}$ class this perturbation
cannot be connected smoothly in the $d-b$ plane to $EC$ or $ZZ$
perturbations since the latter belong to the $G$ class of 
perturbations
that are antisymmetric under the mirror glide operation. Since we decided to
transform all perturbations into the $G$ class by a shift
$d' = d - k$ as explained in ~\S\,\ref{sec241} we have to rewrite the above
$CR$ perturbation (\ref{CRmode}) into the form
\begin{equation}
\delta \Phi_{-101} e^{i(d-k)x + i b y} C_1(z) \; .
\end{equation}
Writing out the CR perturbations in this form one finds again 
that the corresponding eigenvalue is connected to the eigenvalue of the 
lateral shift, only taken at a different $d = k$. 

Since the CR instability does not occur at arbitrarily small $b$ 
one therefore has to test several values 
of $b$ and then apply an interpolation procedure.  

\subsubsection{Skewed varicose instability}
\label{sec414}

This instability is not captured by the simple amplitude 
equations. The SV boundary confines the 
stability balloon on the large-$k$ side. When crossing this boundary the
perturbation tends to replace the original set of rolls 
by a new set with smaller wavenumber. The eigenvalue has its maximum at 
$d \neq 0$ and $b \neq 0$. The SV instability too reduces to a lateral shift 
at $d,b \rightarrow 0$ and can be found at infinitesimal small $b$ and $d$.
To find the SV stability boundary of rolls one has to find the maximum of the
eigenvalue on a line between the $d$-- and $b$--axis near the origin.

\subsubsection{Oscillatory instability}
\label{sec415}

The oscillatory instability is the only instability with complex
eigenvalues at small $r$. The perturbation is most critical at $d = 0$ and 
finite $b$. It is even in $x$. The pair of complex
conjugate eigenvalues undergoes near $b = 0$ a collision and generates two
real eigenvalues. One of these stationary perturbations transforms into the 
lateral shift. The real eigenvalues are negative (if the pattern is stable 
against ZZ perturbations), so the search for the OS instability requires an 
evaluation and interpolation along the $b$--axis as for the CR instability.

\subsection{Stability boundaries of rolls}
\label{sec42}

We saw that the EC, ZZ, CR, SV, and one eigenvalue of the OS instability are
connected to the lateral shift at $b=d=0$. Figure~\ref{bdplane} shows an 
example of the most dangerous eigenvalue in the $(d,b)$--plane. 
Along the $d$--axis the value of $s$ goes down from $s=0$ at the origin. The
pattern is therefore EC stable. The maxima at $d=0$ and $d=k$ show that
it is unstable against ZZ and CR perturbations. An oscillatory instability 
does not occur here for these parameters since
the eigenvalue is always real. There is also no SV instability.
The latter would cause a relative maximum between the $d$-- and $b$--axis.

Besides the perturbation at $d = b = 0$ which describes a lateral
shift in all fields there is another important location in 
wavenumber space that is of relevance for concentration field
perturbations and which thus is specific to binary mixtures. It lies at 
$d = k$, $b = 0$ and describes a change of the mean concentration. This is
most easily understood by transforming this perturbation from the $G$
class into the $\overline{G}$ class where it then occurs at $d = b = 0$.
Here it consists of one single mode $\delta \zeta_{000}$.
This mode is constant in the fluid layer and describes a change 
in the average concentration. But since only derivatives of the 
$\zeta$--field show up in the
balance equations such a mode has no influence so that $s=0$. The reason why
this mode has to be taken into account is that for $d$, $b \neq 0$ it
describes a long--wavelength perturbation
$\delta \zeta_{000} e^{idx + iby}$
that is indeed of physical importance. Because the average 
concentration is fixed, a divergence in
the $\delta \zeta_{000}$--amplitude (as in $\delta \Phi_{001}$ for mean flow)
cannot occur when approaching $b,d = 0$. 
While the eigenvalue of the CR perturbation is connected to the above 
described zero eigenvalue the
CR instability occurs always at finite $b$. Only far in the unstable region
the CR eigenvalue becomes positive at arbitrary small $b$. 
However, we found no roll instability 
to occur directly near this point whereas EC, ZZ, and SV instabilities are 
realized near the origin $d = b = 0$ as modifications of the lateral shift.

Figure~\ref{CRevolution} shows the CR instability boundaries of rolls 
in the full Galerkin model
in a parameter interval where an exchange of stability between rolls and 
squares at onset is predicted in (\cite[Clune \& Knobloch 1991]{CC91}). One 
sees that the curvature of the CR boundary at the critical point
diverges --- as is also predicted by the amplitude equation 
(\ref{ampligleich1} - \ref{ampligleich2}) for $f = 1$ --- 
when this exchange occurs. For the parameters $\sigma =10, \psi=0.01$ of 
figure ~\ref{CRevolution} the exchange occurs at $L=0.2$. Decreasing
$L$ below this value the neutral stability curve 
(dashed line in figure~\ref{CRevolution}) drops further down in $r$ 
(not shown in figure~\ref{CRevolution}b) while the CR instability boundary
moves up in $r$. The $r$-range between these two curves locates stable squares.
In such a situation where squares are stable at onset the amplitude equations
predict rolls to be CR-unstable everywhere while in the full equations rolls 
become stable against CR-perturbations above the solid lines in 
figure~\ref{CRevolution}.

The rolls could still be unstable there against other perturbations but we 
found the minimum of the CR boundary always to be the minimal Rayleigh number
for stable rolls to exist. To know the location, $r^c_{CR}$, of this minimum of
the CR boundary for
several parameters is therefore of interest. We have determined 
$r_{CR}^c$ and the associated wavenumber $k_{CR}^c$ 
of the most dangerous CR perturbation for $\psi = 0.01$ and $\psi = 0.15$ 
and presented the results in figures~\ref{crit0.01} and~\ref{crit0.15}, 
respectively. They show that rolls are CR unstable at onset, $r_c$ 
(dashed lined in figures~\ref{crit0.01}a and ~\ref{crit0.15}a), for small $L$
and large $\sigma$. However, for larger $L$ 
being typical for gas mixtures, rolls are stable at the critical point. 

At $\psi=0.15$ the minimum $r^c_{CR}$ of the CR boundary and its wavenumber 
location $k^c_{CR}$ strongly increase for small $\sigma$ and $L$ --- cf. the
small-$L$ variation of the respective curves at $\sigma=0.2$ in figures
\ref{crit0.15}a,b. $r^c_{CR}$ seems to diverge here, but a detailed inspection 
shows that this is not true. The two branches of the CR boundary meet again at 
higher $r$ near the apparent divergence. They limit in the $k-r$ plane an
oval region of CR--stable rolls from below and also {\em above}. 
(The upper part of this region can be unstable against OS--perturbations, 
though.) By reducing $L$ $r^c_{CR}$ remains finite but the oval region gets 
smaller until the region of stable rolls vanishes.
The experimental observation of this behaviour might be
difficult because it occurs in a region of the parameter space that is not
accessible by ordinary fluid mixtures.

We did also investigate the behaviour of $b^c_{CR}$, the wavenumber of the 
critical perturbation at $(k^c_{CR},r^c_{CR})$.
Only in the Rayleigh region the values of $k^c_{CR}$
and $b^c_{CR}$ are near $k_c = 3.117$. But in general $k^c_{CR} \neq
b^c_{CR}$. The {\em linear} analysis gives thus a hint for the existence of
patterns with $k_x \neq k_y$ in the Soret region.  

We have calculated all stability boundaries at small $r$ for different values 
of $L$, $\psi$, and $\sigma$. Concerning the stability behaviour of the roll 
structures one sees that in the Rayleigh region, $r \gtrsim 1$, where the 
concentration field is nearly uniform the 
binary mixture behaves like a pure fluid. And the transition between Soret
and Rayleigh region is very sharp at small $L$ and $\psi$. An example for such 
a behaviour is given in figure~\ref{exampleballoon}. Here only the EC, CR,
and ZZ boundaries are of importance. In the Rayleigh region $r\gtrsim 1$ 
of figure~\ref{exampleballoon} the CR, ZZ, and EC boundaries of the mixture
(full lines with circles, triangles, and squares, respectively) are lying close 
to the corresponding boundaries of the pure fluid (long-dashed lines). Note 
in particular the vase-like form of the EC boundary $r_{EC}(k)$ and the dent 
in the ZZ boundary
$r_{ZZ}(k)$: Close to onset ($r_c \simeq 0.6, k_c \simeq 2.6$ in 
figure~\ref{exampleballoon}), i.e., in the Soret regime $r_{EC}(k)$ opens up 
parabolically and $r_{ZZ}(k)$ comes out of the critical point linearly with
{\em positive} slope. However, in the crossover range $r \sim 1$ between
Soret and Rayleigh regime the curve $r_{EC}(k)$ pinches inwards and develops 
a waist such as to follow in the 
Rayleigh regime the parabolic shape of the EC curve of the pure fluid that 
starts out at $k_c^0 \simeq 3.1, r_c^0=1$. Similarly $r_{ZZ}(k)$ bends in the 
crossover range towards small $k$ to follow then the ZZ boundary of the pure 
fluid that shows {\em negative} slope.

Figures~\ref{balloon0.1}--\ref{balloon10} show that this sharp 
transition between Soret and
Rayleigh region that causes the vase-like structure of the EC boundary and 
the sharp bend of the ZZ boundary does not occur when $L$ is greater. For
$L=0.01$ and $\psi \geq 0.08$ the critical point lies at $k = 0$. In this
case the amplitude equations are not applicable. We found the ZZ boundary and 
the left EC branch to go to $r = \infty$ at small $k$. But the right EC 
branch does still meet the critical point. 

At low Prandtl numbers the roll solutions have a finite region of stability but
their basin of the attraction seems to be very small since typically spiral
defect chaos is observed here in experiments (\cite[Liu \& Ahlers 1996]{LA96}). 
But it was shown that rolls could also be observed with special experimental 
procedures (\cite[Cakmur \etal 1997]{CEPB97}).

We tried to find laws that connect the position of the boundaries and the
fixed point solutions. We found no easy connection between these values for CR, 
ZZ and SV. But the OS boundary for fixed $\sigma$ seems to be independent of
$L$ and $\psi$ in the plane of $k$ and the convection amplitude 
$w_{101}$ instead of in the $k$--$r$  plane (figure~\ref{OSscale}). 

For the EC boundary a more complicated 
procedure is needed. We define an effective control parameter by linear
interpolation of the values for the convection amplitudes 
\begin{equation}
\frac{\partial w_{101}^2}{ \partial r} \epsilon_{eff}(r) = w_{101}^2 \;\; , 
\label{epsilon_eff} 
\end{equation}  
and an effective reduced Rayleigh number
\begin{equation}
r_{eff} = \epsilon_{eff} + r^0_{stab}(k)\;\; . 
\label{r_eff} 
\end{equation} 
Within the amplitude equations it is just 
$\epsilon_{eff} = \epsilon$. Plotting the EC boundary in the $k - r_{eff}$ 
plane instead of in the $k - r$ plane shows in the Rayleigh region only a 
dependence on $\sigma$ but not on $\psi$ and $L$, c.f. figure~\ref{ECscale}. 
However, this procedure does not hold in the Soret region. 

\subsection{Stability boundaries of squares}
\label{sec43}

Performing the stability analysis of squares we had to restrict ourselves to the
case $d=b=0$ where the perturbations separate into different symmetry
classes. An analytical investigation of long wavelength perturbations of 
squares near the critical point can be found in (\cite[Hoyle 1993]{H93}). 
Because both squares and rolls can be described as even in $x$ and
$y$ and as mirror glide antisymmetric, one expects a perturbation that
destabilises the squares and favours the rolls to fulfill these symmetries, too.
However such a perturbation should break the symmetry $x \leftrightarrow y$. 
We actually always found the most critical perturbation to fall into this
symmetry class. Other perturbations that break the mirror symmetry in 
$x$-- or $y$--direction are less critical. 

Figure \ref{squareballoon} shows typical examples for the stability
region of squares. The left and right side of the stability boundaries should 
not be taken too serious --- presumably square structures are destabilised
earlier by instabilities with finite $b$ or $d$ that tune the wavenumber and 
that are not considered here. 

Even if squares are stable at onset they always lose  their stability
against a roll pattern at higher $r$. Furthermore, for certain parameters 
there does also exist a band of
$r$--values where neither squares nor rolls are stable. Within this band
three--dimensional CRs that break the $x \leftrightarrow y$ symmetry
can be stable. For certain parameters we found stationary CRs only. For others
oscillating CRs structures appeared as well. We have studied these
structures in more detail in (\cite[Jung, Huke \& L{\"u}cke 1998]{JHL98}).

\section{Conclusion}
\label{conclude}

We investigated roll, crossroll, and square convection in binary mixtures 
for a wide range of parameter combinations using a multi--mode Galerkin method. 
All these patterns are realized as stable convection structures somewhere in
parameter space. 
The bifurcation behaviour of rolls and squares can be modelled near onset by 
amplitude equations, which, however lose their applicability in the Rayleigh 
region. Moreover, the CR solution that connects the roll and the 
square branch at Rayleigh numbers $r \approx 1$ is absent in
a simple ansatz of cubic amplitude equations. 

We compared the Nusselt number and also the mixing parameter of the full 
numerical solutions for rolls and squares. We found these global properties
of the convective states to be approximatively the same for these patterns 
despite the qualitatively different structures of 
the concentration fields of squares and rolls: Rolls show a strong boundary 
layer behaviour at small $L$ and high $r$, so that compared to pure fluids 
much more modes are needed to describe them. Squares, on the other hand,
show a less pronounced boundary layer behaviour making it easier to 
determine these three--dimensional solutions numerically. 

In the main part of the paper 
we investigated the linear stability of rolls and squares. To that end we
performed a full and unrestricted stability analysis for rolls using 
arbitrary perturbations and in addition a 
stability analysis of squares that uses the periodic boundary conditions of 
the squares also for the perturbations. The result is that squares are 
stable in the Soret region if $L$ is sufficiently small. But they always 
lose their stability at higher $r$. Rolls are stable at onset at higher $L$ and 
in the Rayleigh region where squares are unstable. Typically, however, there 
is a finite interval of $r$--values in between where neither rolls nor squares 
are stable. In this interval either stationary or oscillatory CR patterns are 
observed. 

The analysis of the rolls shows that in the explored parameter range
only the basic mechanisms of instability 
occur that are already known from the pure fluid, namely the Eckhaus, zigzag, 
crossroll, oscillatory, and skewed varicose mechanism. When the Soret region is 
small, the stability balloon of the mixtures resembles the Busse balloon for
the pure fluid. However, at small $L$
when the Soret region is large the situation is different. The fixed point 
solutions show a sharp transition between the two regimes. The convection 
amplitudes are very small in the Soret regime. But near $r =1$ they increase 
strongly and e.~g.\ the Nusselt number becomes comparable to that of the pure 
fluid convection. The stability boundaries of roll convection show a similar 
transition here. In the Rayleigh region the boundaries are close to the 
boundaries of the pure fluid. But upon reducing the Rayleigh number
the boundaries begin to deviate from their pure fluid 
counterparts. Near onset they finally agree with the predictions of the 
amplitude equations for the mixtures.

The EC boundary gives a typical example for this behaviour. For high $L$ it has 
just a normal parabolic shape. But for small $L$ one observes a qualitatively
different behaviour. When approaching $r = 1$ from above the two branches do
not meet near $r=1$ as in the pure fluid but begin to separate again until $r$
gets small enough for the amplitude equations to become valid. A typical 
vase--like shape results.

A detailed inspection of the point where the rolls become unstable at onset 
shows that the CR boundary is responsible for this loss of stability. As long 
as the rolls are stable at onset, the CR boundary touches the critical point, 
as predicted by the amplitude equations.  At a certain point the 
amplitude equations show a global loss of stability against CR perturbations. 
In the full Galerkin analysis this loss of stability does not occur. The 
CR boundary disconnects from the neutral curve and the rolls 
remain stable not at onset but at higher $r$.

Both the amplitude equations and the full Galerkin expansion show that
square structures get stable at onset when roll structures lose stability
there and vice versa. The stability domain of squares lies mainly in the Soret 
region  and it is separated from the region of stable rolls at higher $r$ by 
the region of
CR structures. The fact that square patterns lose their stability already at 
relatively small $r$ was another favourable property which delimited the 
requirements for a numerical analysis of these patterns.

The perturbations against which square solutions are unstable show those 
symmetries that squares, rolls, and CRs have in common. But they
break the $x \leftrightarrow y$ symmetry, which exists only for
squares. The resulting stability boundary delimits the 
region of stable squares in the $k$--$r$ plane not only on top but also at 
the sides at smaller and larger wave numbers. It should be noted that our
analysis of squares does not cover instabilities that retune the actual
wavenumber of the considered pattern and which might occur earlier than the 
wave number preserving instabilities investigated here.

\begin{acknowledgments}
This work was supported by the Deutsche Forschungsgemeinschaft.
\end{acknowledgments}

\clearpage
\begin{figure}
\caption{Flow intensity of stationary squares and rolls.
Shown is the contribution $k^2 (|\Phi_{101}|^2+|\Phi_{011}|^2)$
from the leading modes to $w^2$ versus reduced Rayleigh number.
Parameters are $\sigma = 1$, $L = 0.01$ (a), and $\psi = 0.08$ (b).
The thin curves denote the roll solution in pure fluids ($\psi=0$).}
\vspace{1cm}
\label{flow_intensity}
\end{figure}
\begin{figure}
\caption{ Representative bifurcation diagram of stationary patterns.
The two most important modes of the velocity field are plotted
versus $r$. Squares, circles, and triangles denote the square, roll, 
and CR solutions, respectively. Upwards (downwards) pointing 
triangles refer to, say, $|w_{101}^{CR}|^2$ ($|w_{011}^{CR}|^2$) for
CRs with dominant 101 mode.
Closed (open) symbols denote stable (unstable) structures.
In the interval $1.11 \lesssim r \lesssim 1.15$ the 
stationary solutions are unstable but an oscillatory solution is stable
(\cite[Jung \etal 1998]{JHL98}). 
Parameters are $L = 0.01$, $\sigma = 10$, $\psi = 0.15$, and $k = k_c^0$.  
To get a consistent bifurcation diagram all solutions had
to be calculated with the same truncation level $N_2 = 16$ (see text). 
However, this is not sufficient to describe rolls quantitatively.}
\vspace{1cm}
\label{bifdiagram}
\end{figure}
\begin{figure}
\caption{Structural properties of square (a-d) and roll (e,f) convection
for representative liquid parameters ($L = 0.01$, $\sigma = 10$, left column)
and gas parameters ($L = \sigma = 1$, right column) at $r=1, \psi=0.15$.
In (a,b) the concentration distribution of squares at mid height, $z=0$, 
is shown. In (c-f) we show the concentration distribution in
a vertical cross section at $y=0$. The largest vertical upflow is at
$x=y=0$.} 
\vspace{1cm}
\label{shadowgraph}
\end{figure}
\begin{figure}
  \caption{Mixing parameter $M$ (\ref{M}) versus $r$ for rolls (circles) 
and squares (squares). Parameters are $k = 3.117$,
$\sigma=1$, and $L=0.1$(a), $\psi=0.08$(b).}
\vspace{1cm}
\label{mixparam}
\end{figure}
\begin{figure}
  \caption{Variation of a representative eigenvalue $s$ as a function 
           of perturbation wavenumbers $d, b$. The positive 
           maximum at $d = 0$ and $b \approx 1.5$ ($d = k$ and $b \approx 3.0$)
           implies ZZ (CR) instability. Parameters are $r = 1$, 
           $k = 2.7$, $\sigma = 10$, $\psi = 0.08$, and $L = 0.1$.}
\vspace{1cm}
\label{bdplane}
\end{figure}
\begin{figure}
  \caption{CR instability boundaries (solid lines) of rolls 
           in the Soret region for $\sigma = 10$, 
           $\psi = 0.01$ and several values of $L$. Rolls are stable against 
           CR-perturbations above the solid lines. (a): For $L \geq 0.2$ the 
           CR boundaries touch the neutral curve (dahed lines) in the
           critical point. (b): For $L < 0.2$ the neutral curve goes further 
           down (not shown) and is disconnected from the CR boundaries.
           Then rolls are not stable at the critical point anymore.}
\vspace{1cm}
\label{CRevolution}
\end{figure}
\begin{figure}
  \caption{(a): The minimum of the CR--boundary $r^c_{CR}$ denoting the 
            smallest $r$ where CR--stable 
            rolls exist as a function of $L$ for several values of 
            $\sigma$ and $\psi = 0.01$. (b): The corresponding wavenumbers 
            $k^c_{CR}$.}
\vspace{1cm}
\label{crit0.01}
\end{figure}
\begin{figure}
  \caption{Same as figure~\ref{crit0.01}, but for $\psi = 0.15$.}
\vspace{1cm}
\label{crit0.15}
\end{figure}
\begin{figure}
  \caption{Stability boundaries of rolls. Solid lines with open symbols refer 
           to a mixture with $\sigma=7$, $\psi=0.01$, and $L=0.025$. The 
           corresponding boundaries in a pure fluid with $\sigma=7$ 
           are the long dashed curves. Dotted curves labelled
           EC(AE) and ZZ(AE) are the predictions of the cubic amplitude 
           equations for the EC and ZZ boundaries of the mixture, respectively.}
\vspace{1cm}
\label{exampleballoon}
\end{figure}
\begin{figure}
  \caption{Crossections of the stability balloon of rolls in the $k-r$ plane 
           at $\sigma=0.1$. Open circles: CR, open squares: EC, open 
           triangles: ZZ, filled circles: SV, filled squares: OS. S denotes 
           the region of stable rolls. For $\psi=0.15, L=0.01$ there are no 
           stable rolls.}
\vspace{1cm}
\label{balloon0.1}
\end{figure}
\begin{figure}
  \caption{Crossections of the stability balloon of rolls in the $k-r$ 
           plane at $\sigma=1$. Symbols have  the same meaning as in 
           figure~\ref{balloon0.1}.}
\vspace{1cm}
\label{balloon1}
\end{figure}
\begin{figure}
  \caption{Crossections of the stability balloon of rolls in the $k-r$ 
           plane at $\sigma=10$. Symbols have  the same meaning as in 
           figure~\ref{balloon0.1}.} 
\vspace{1cm}
\label{balloon10}
\end{figure}
\begin{figure}
  \caption{(a): OS boundaries $r_{OS}(k)$ for the nine parameter combinations 
           of figure~\ref{balloon0.1} versus $k$. (b): Amplitude square
           $|w_{101}|^2(k)$ of the leading velocity mode of the roll patterns
           at the OS boundaries shown in (a). Thick lines in (a) and (b) 
           refer to the pure fluid at the same $\sigma = 0.1$.}
\vspace{1cm}
\label{OSscale}
\end{figure}
\begin{figure}
  \caption{(a): EC boundaries $r_{EC}(k)$ for the 27 parameter combinations 
           of figures~\ref{balloon0.1} -- \ref{balloon10} versus $k$. 
           (b): EC boundaries of (a) in the $k-r_{eff}$ 
           plane defined by (\ref{epsilon_eff}, \ref{r_eff}). Only data 
           $r_{EC}(k) > r^0_{stab}(k)$ in the Rayleigh region
           are shown. The location of the EC boundary depends only slightly
           on $\sigma$ for $\sigma > 1$. Therefore, only two different 
           boundaries appear in (b). The inset shows the definition of 
           $\epsilon_{eff}$.}
\vspace{1cm}
\label{ECscale}
\end{figure}
\begin{figure}
  \caption{Crossections of the stability balloon of squares in the $k-r$ plane  
           obtained at $\sigma=0.1$ (open squares) and $\sigma=10$ 
           (open circles) from a restricted stability analysis as 
           explained in the text. S denotes the region of stable squares.
           For $\sigma=0.1$, $L=0.1$, and $\psi=0.01$ there are no stable 
	   squares.}
\vspace{1cm}
\label{squareballoon}
\end{figure}

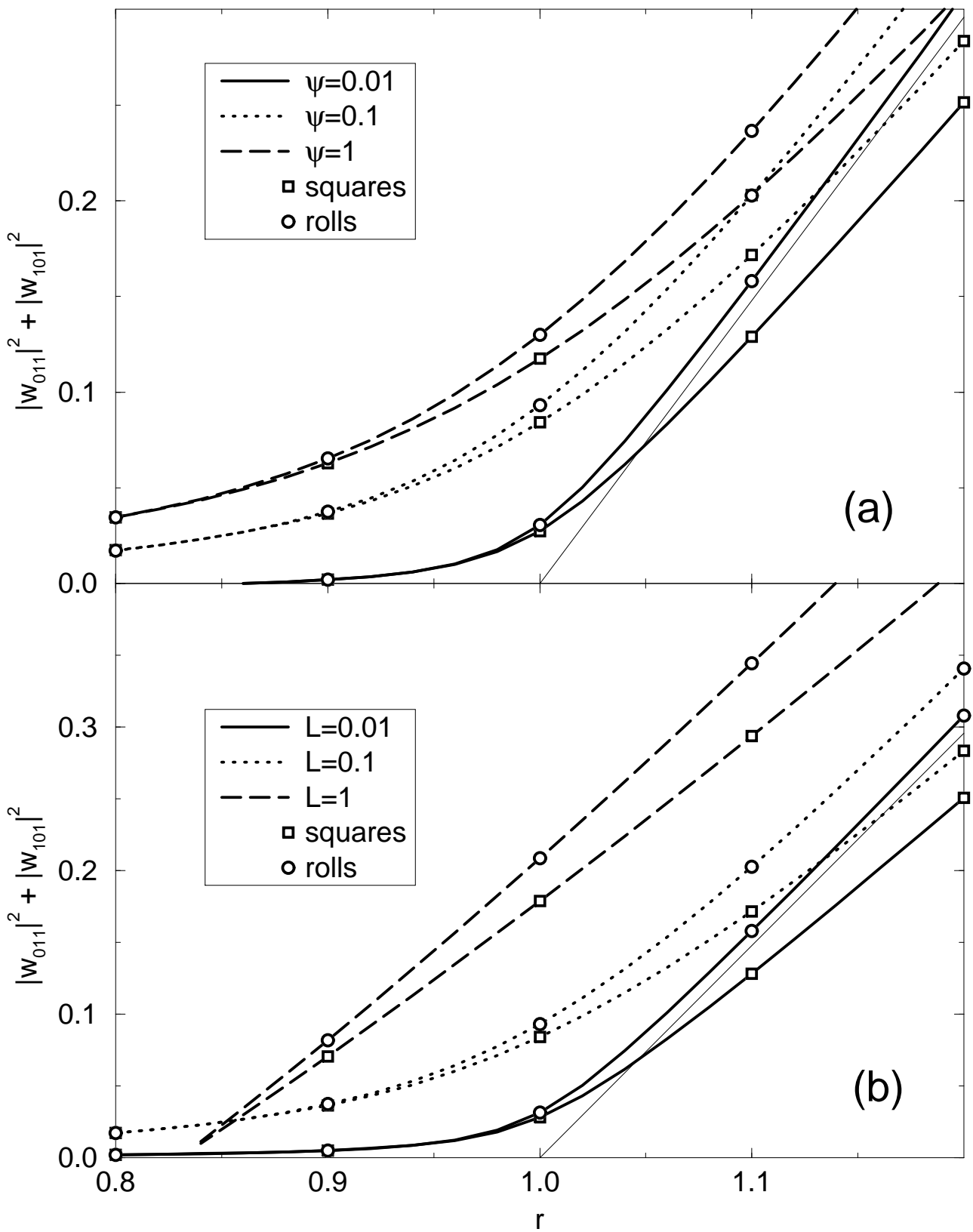
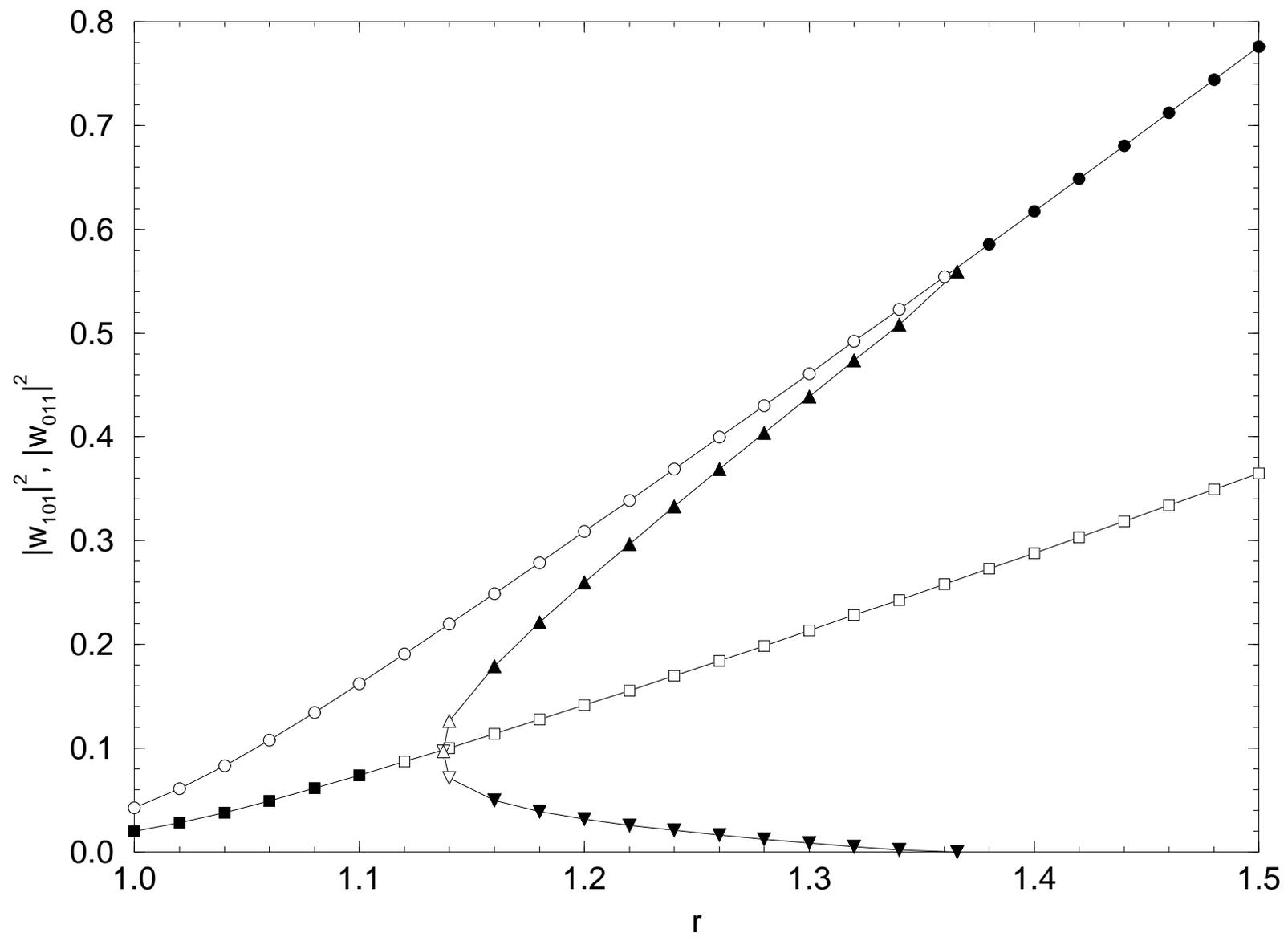
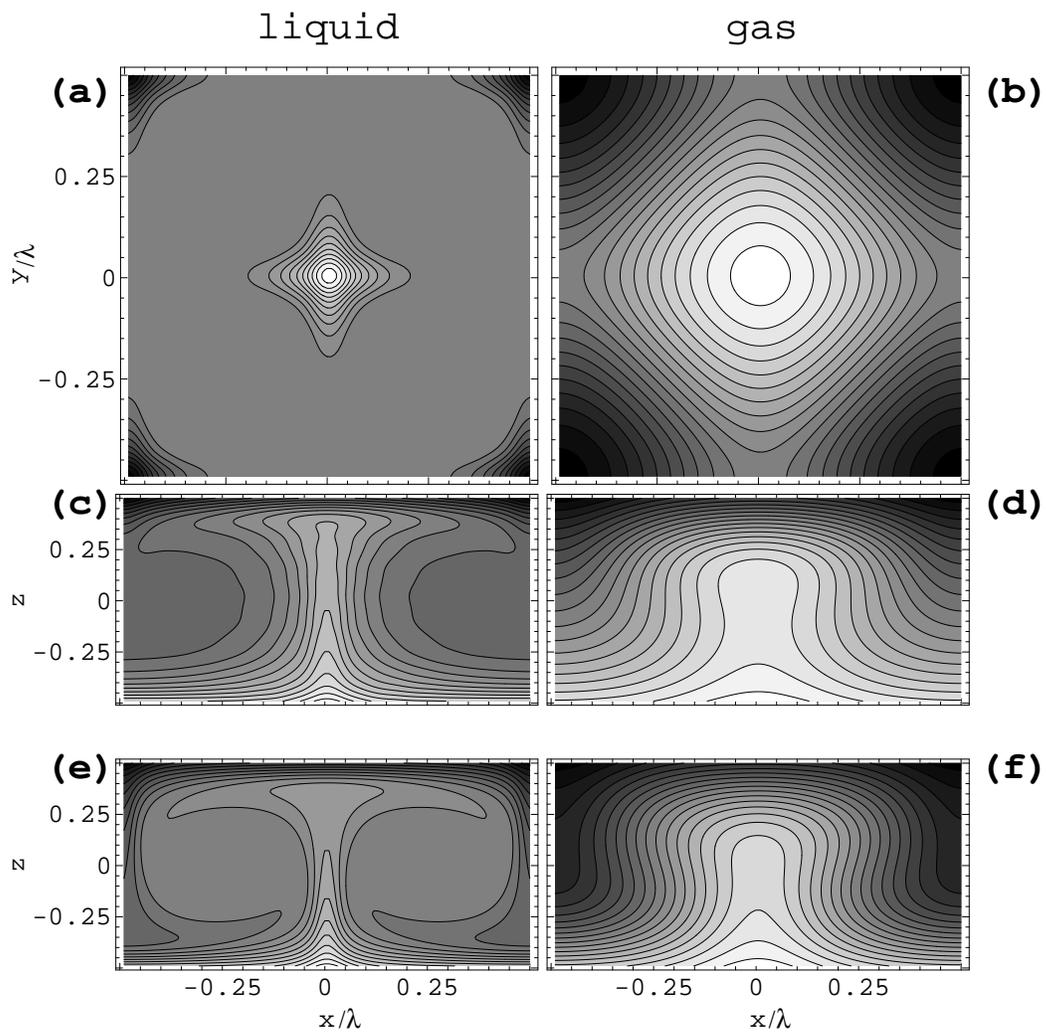
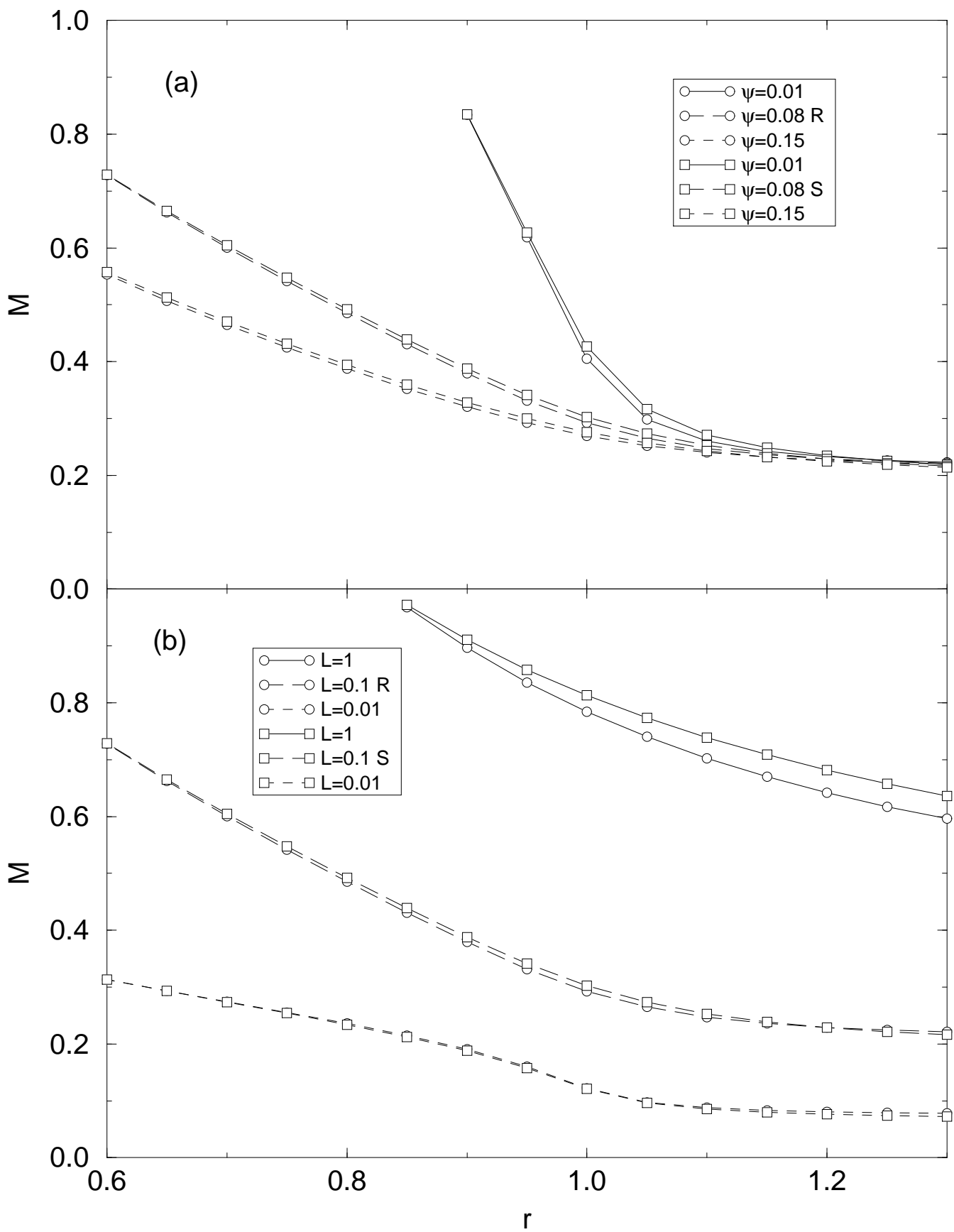
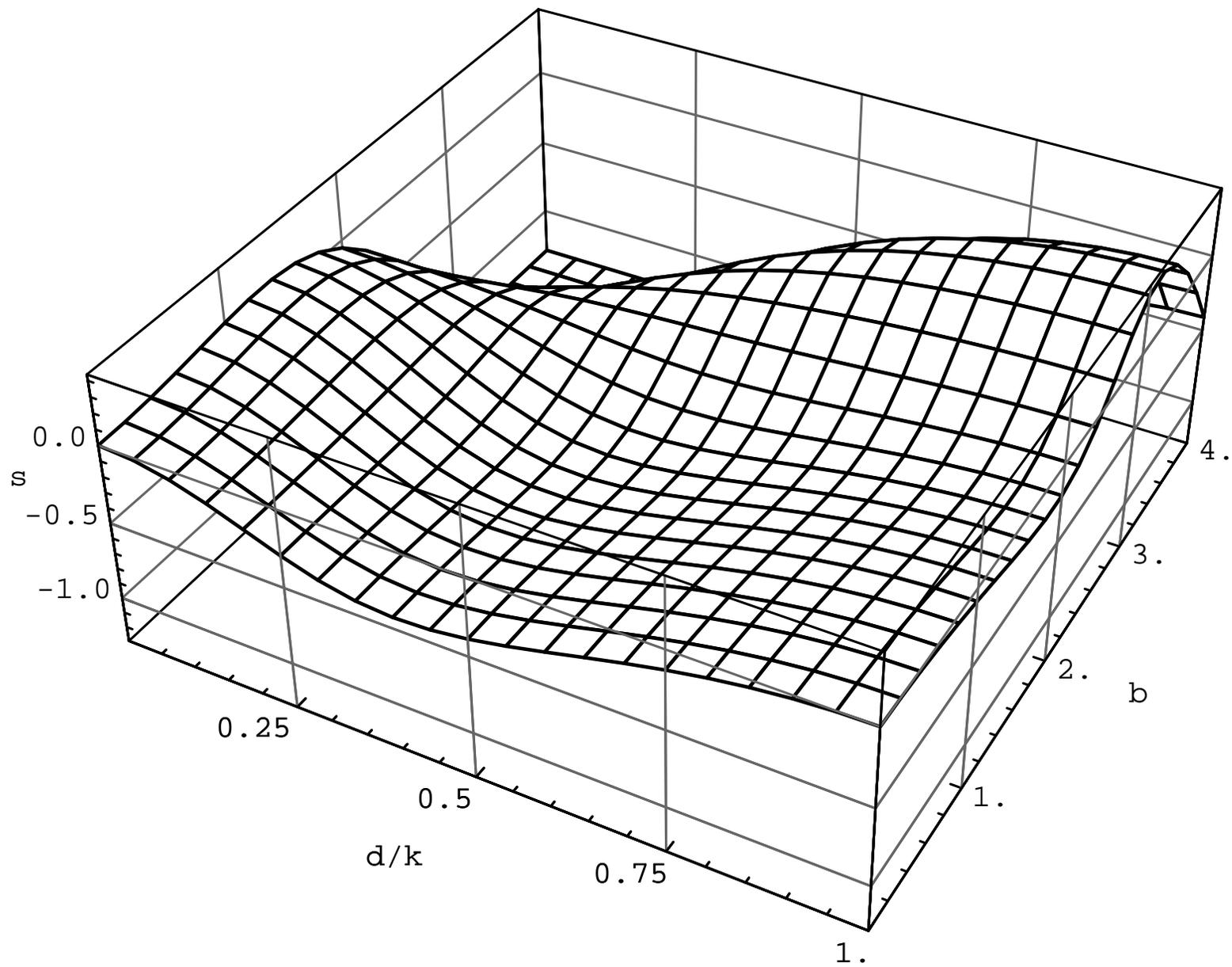
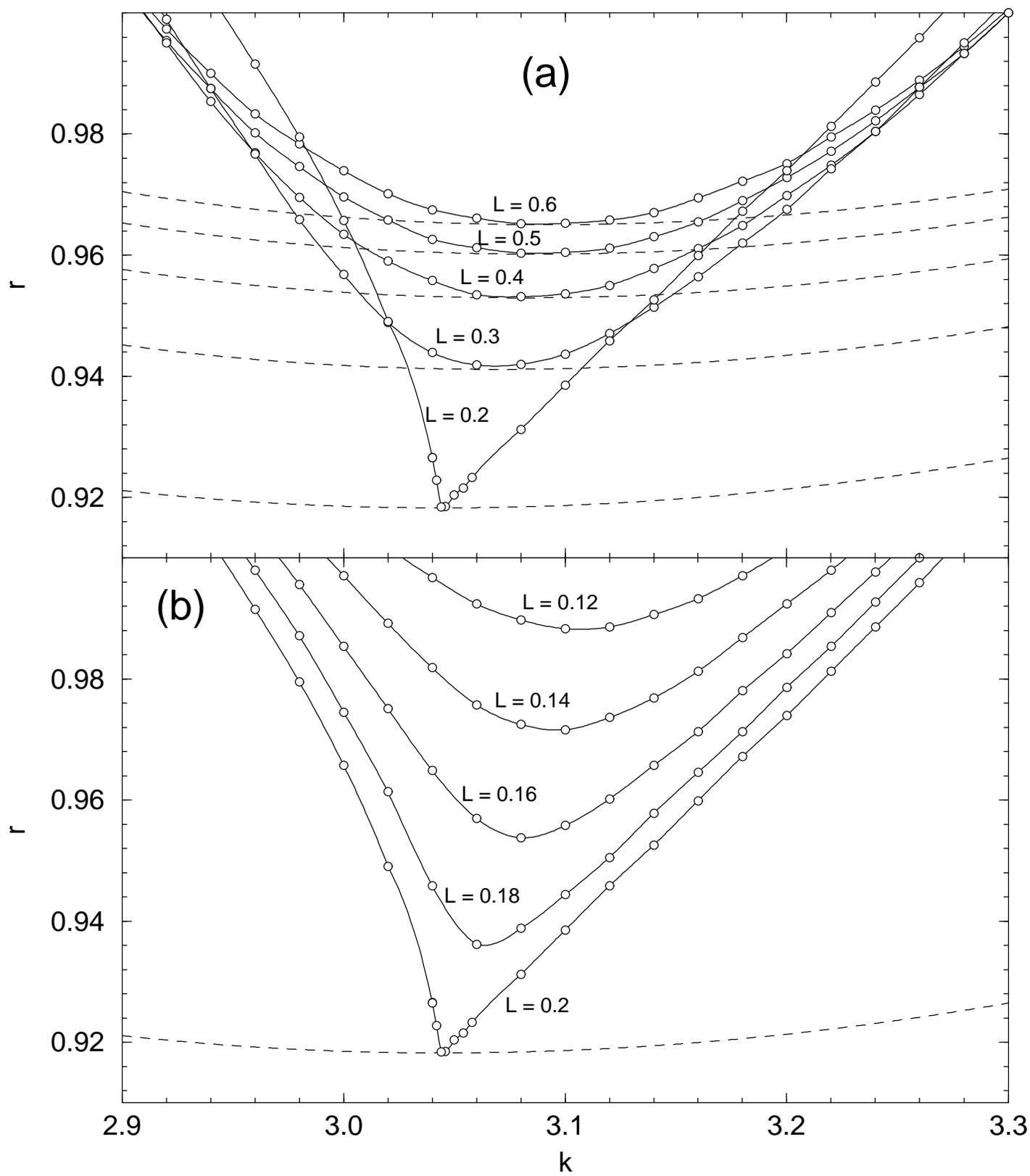
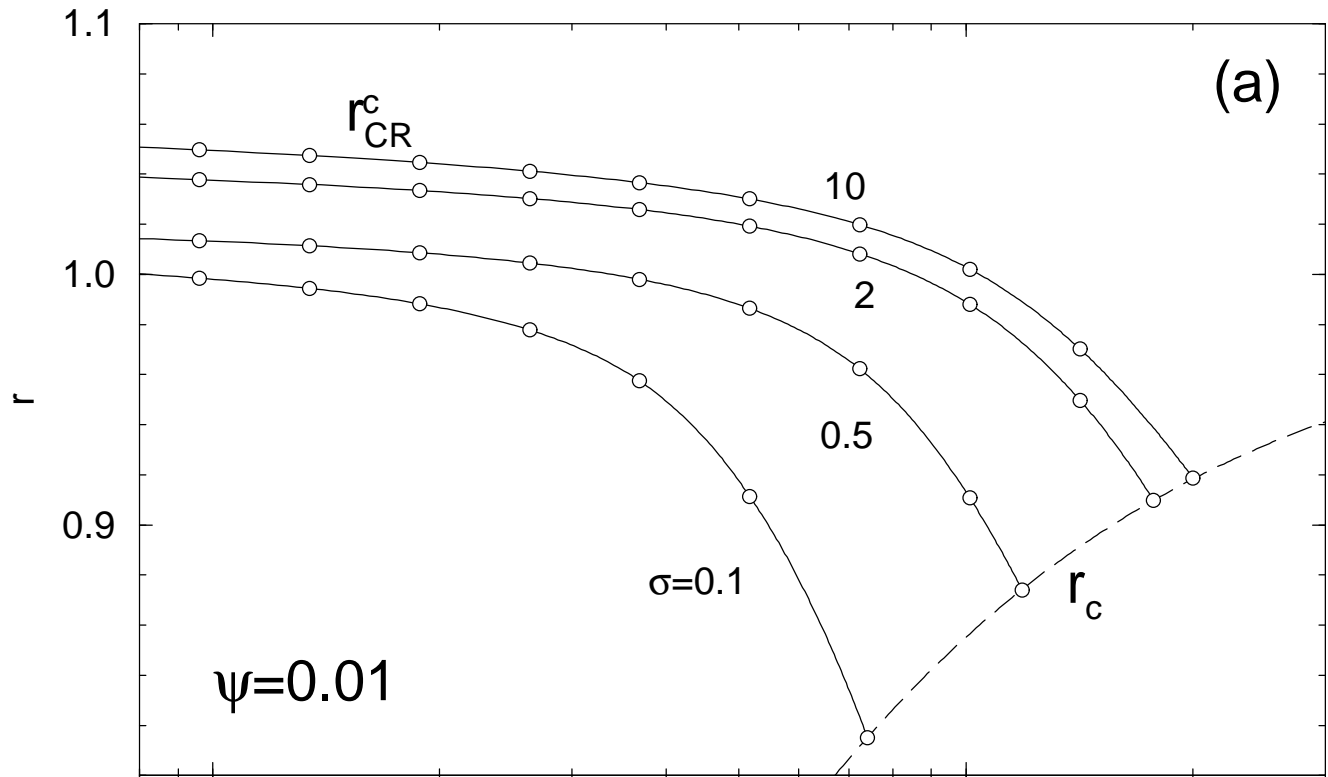
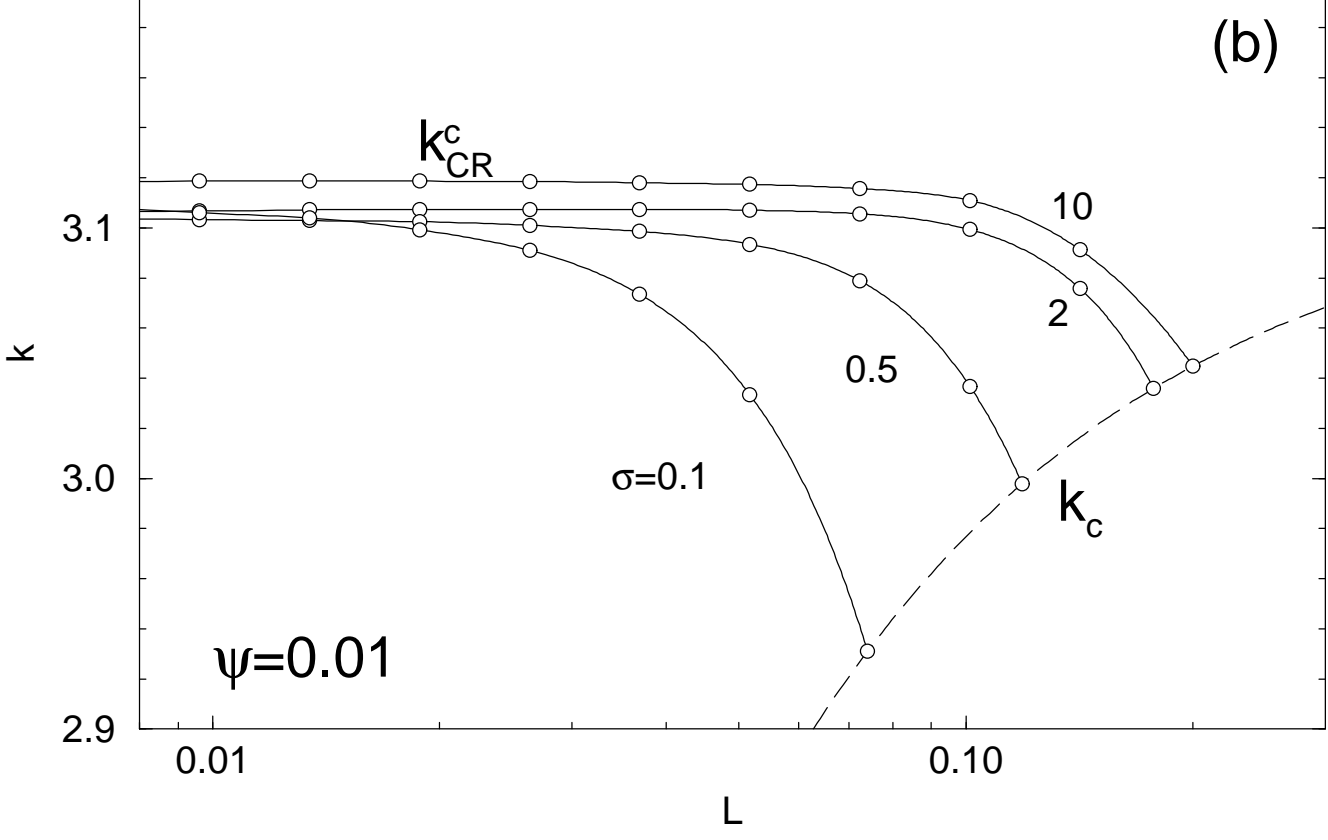
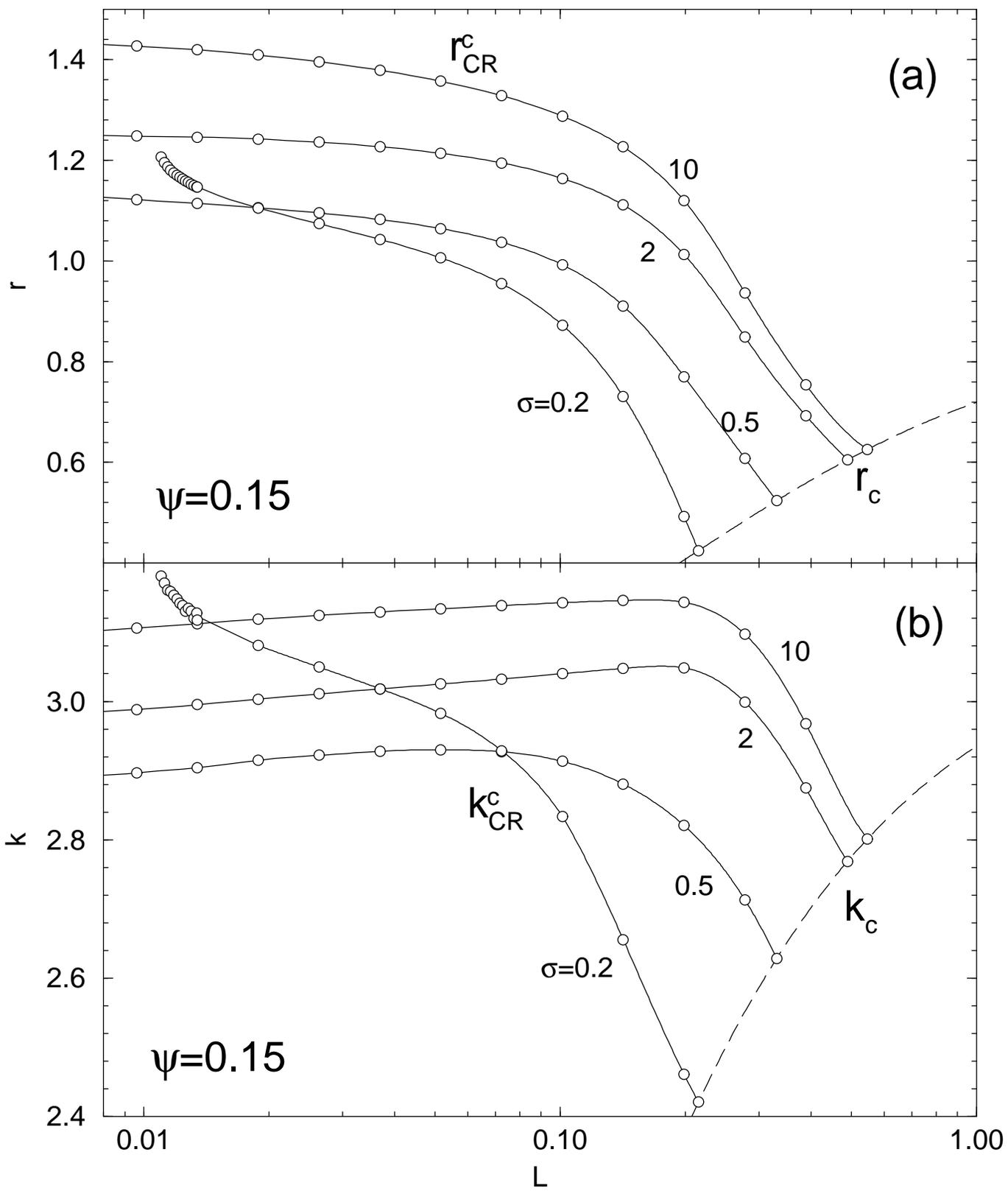
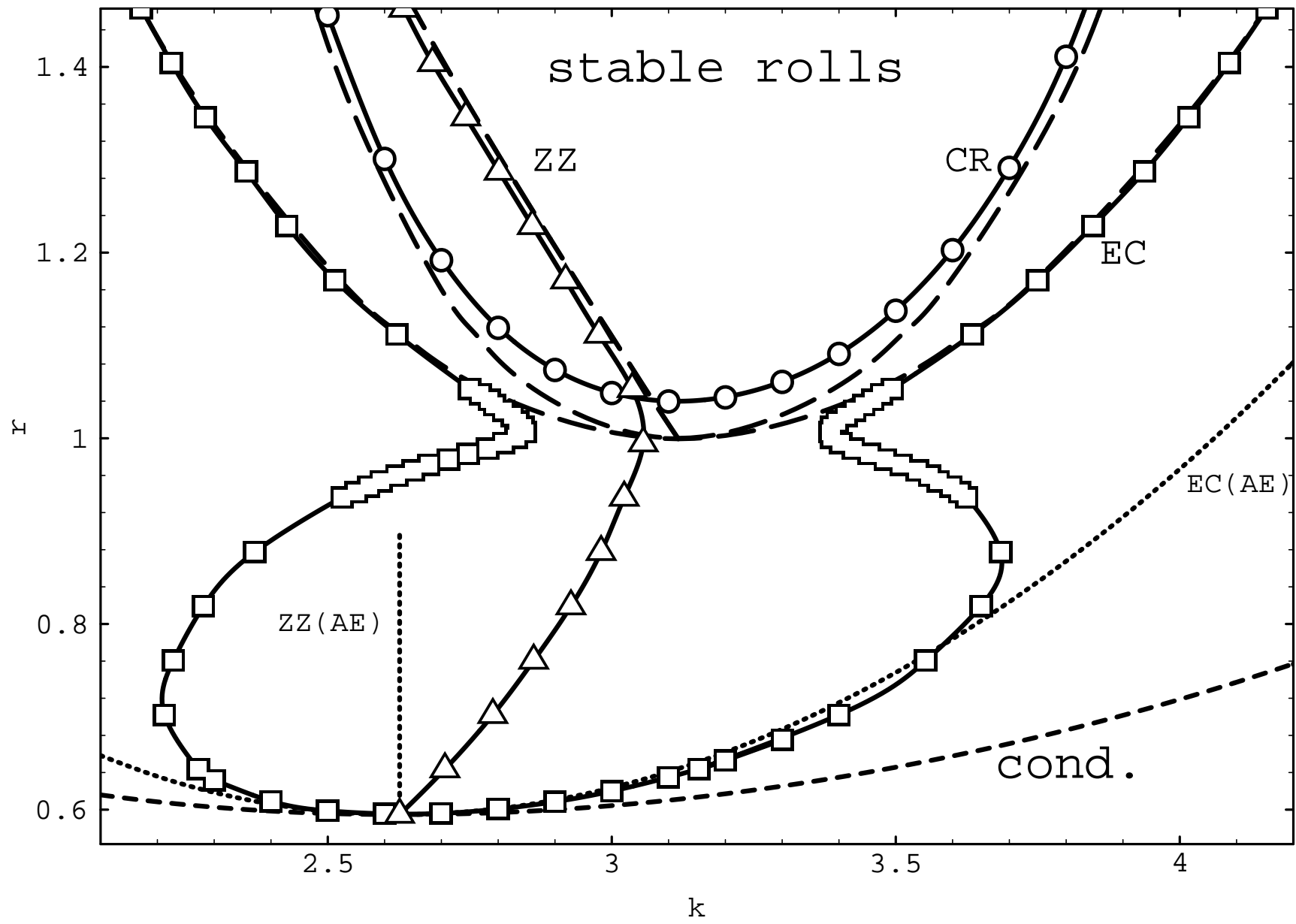
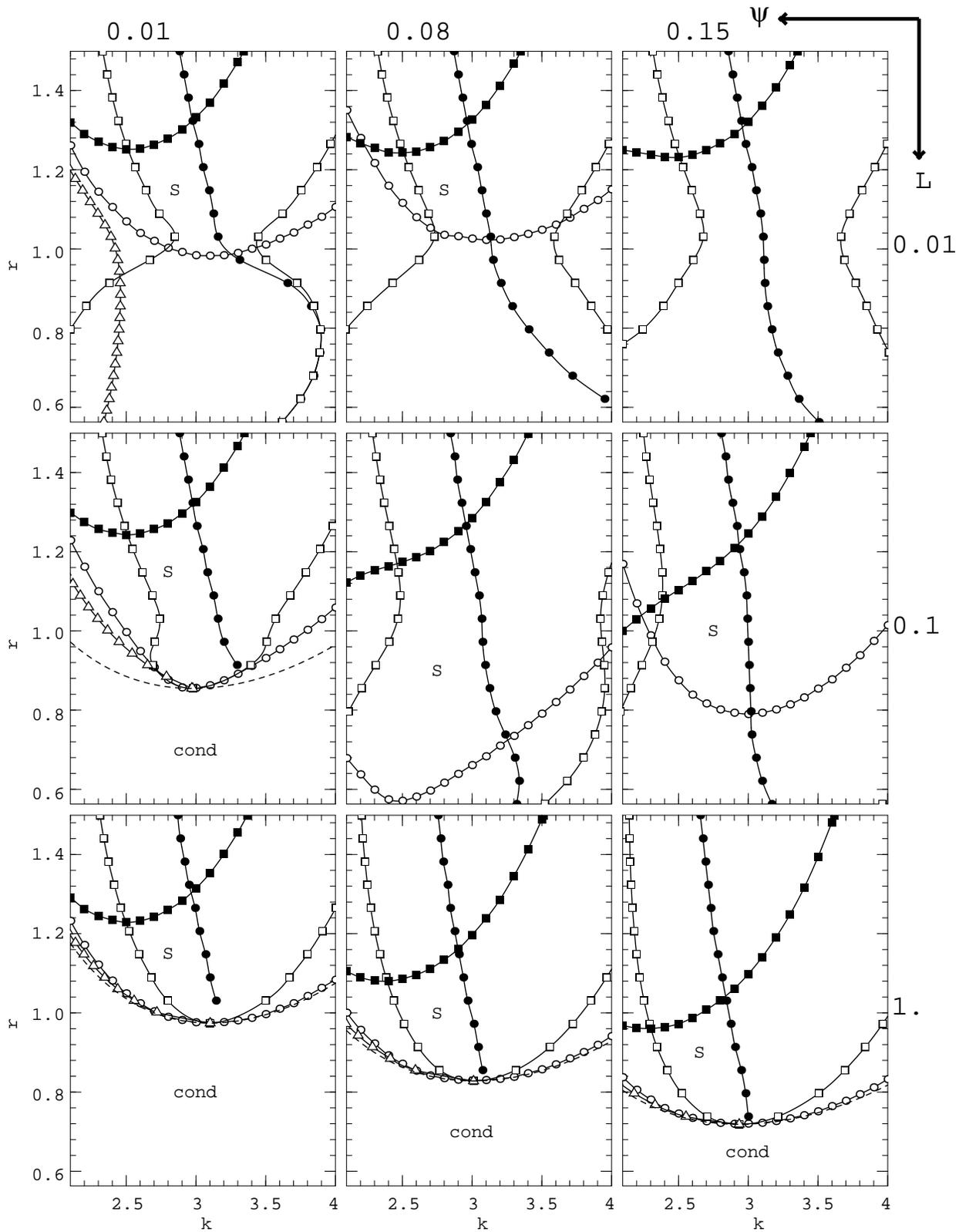
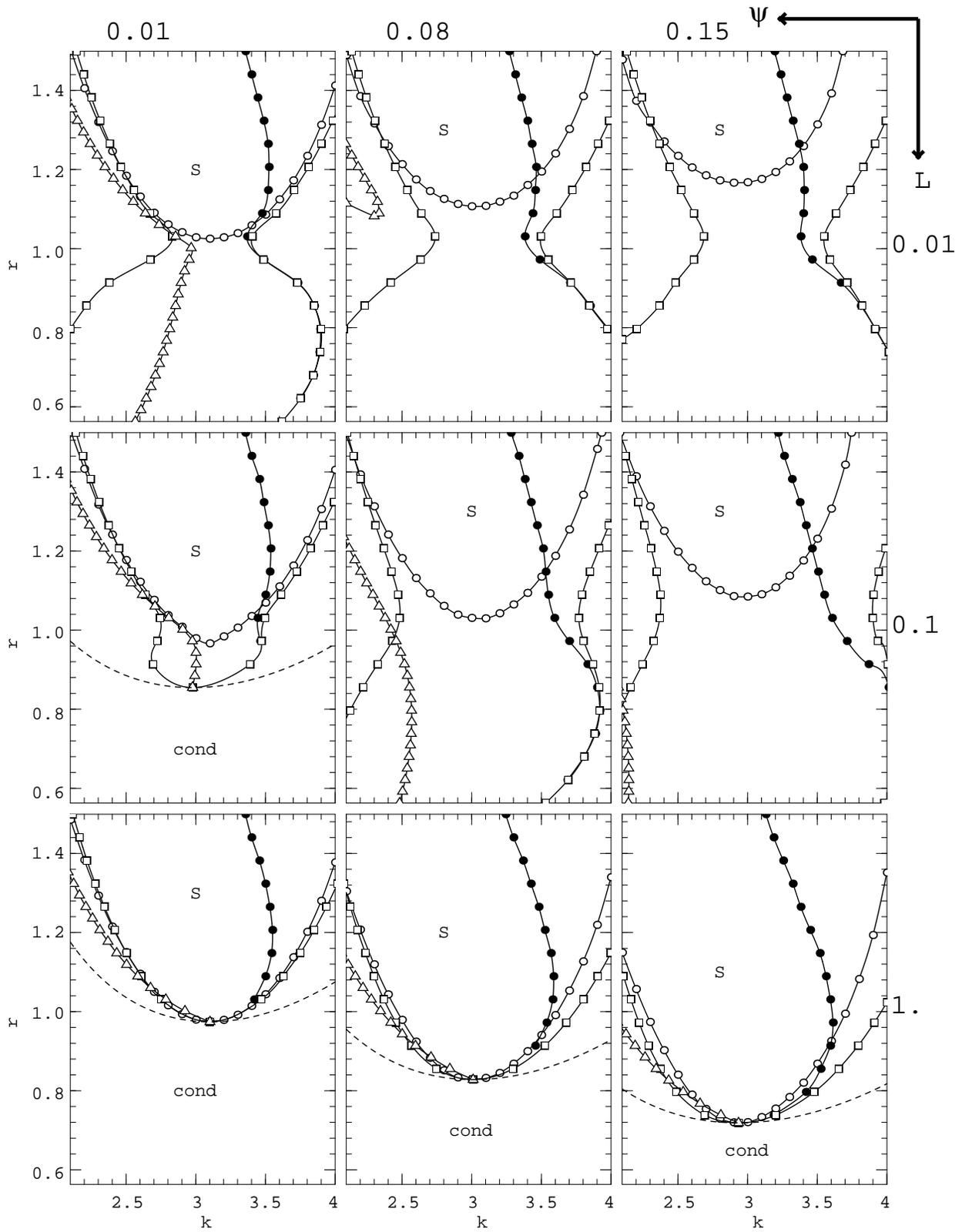
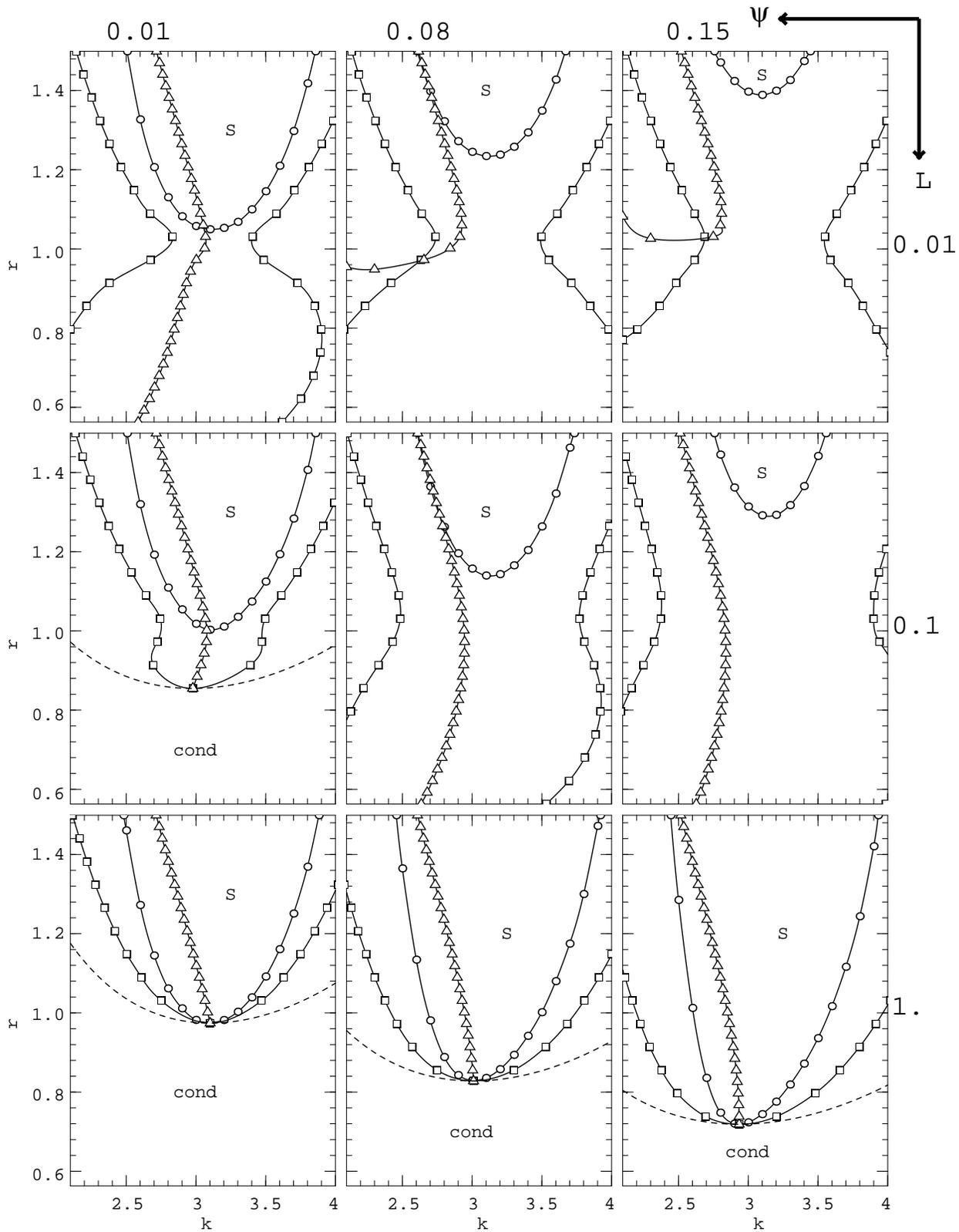
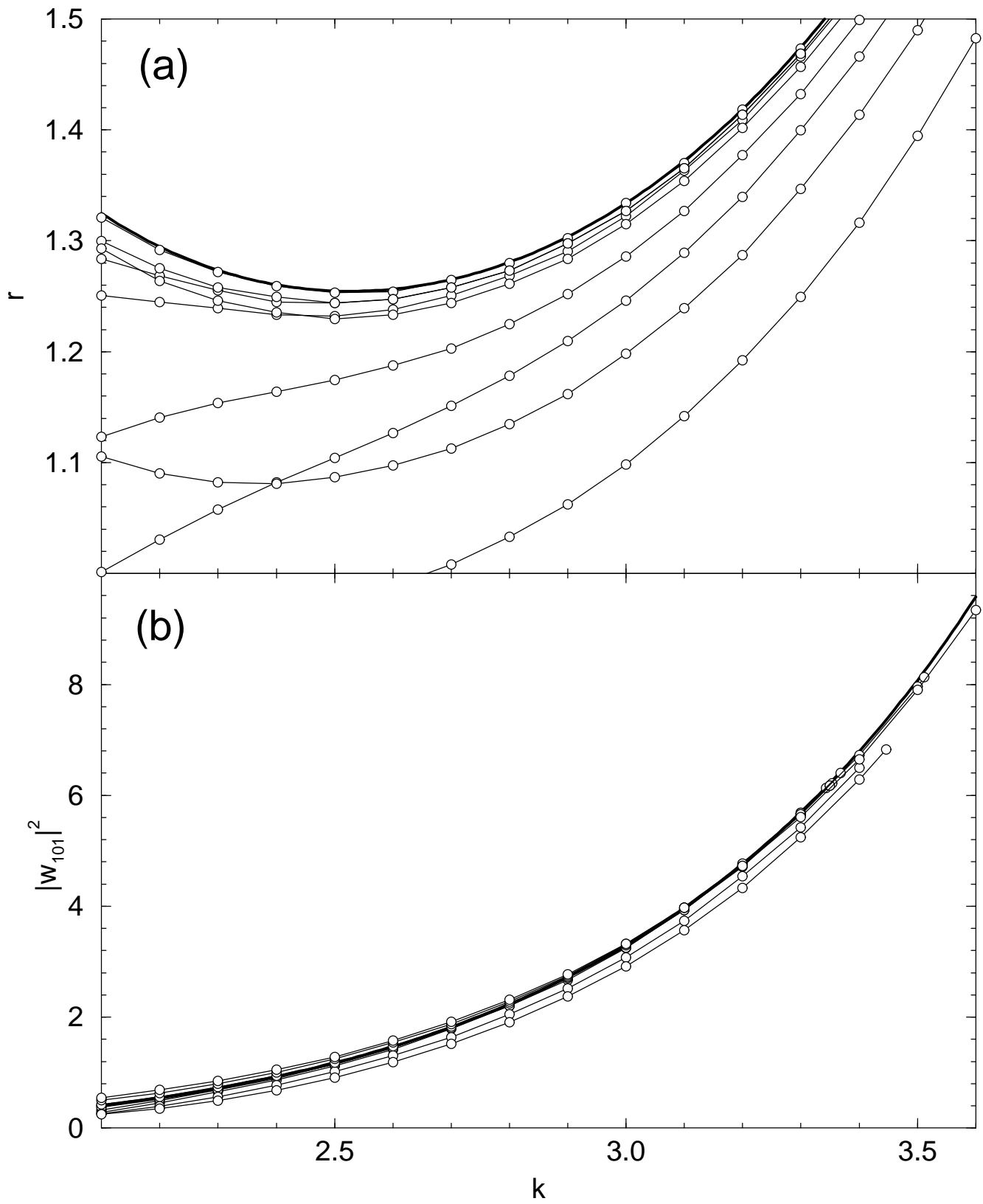
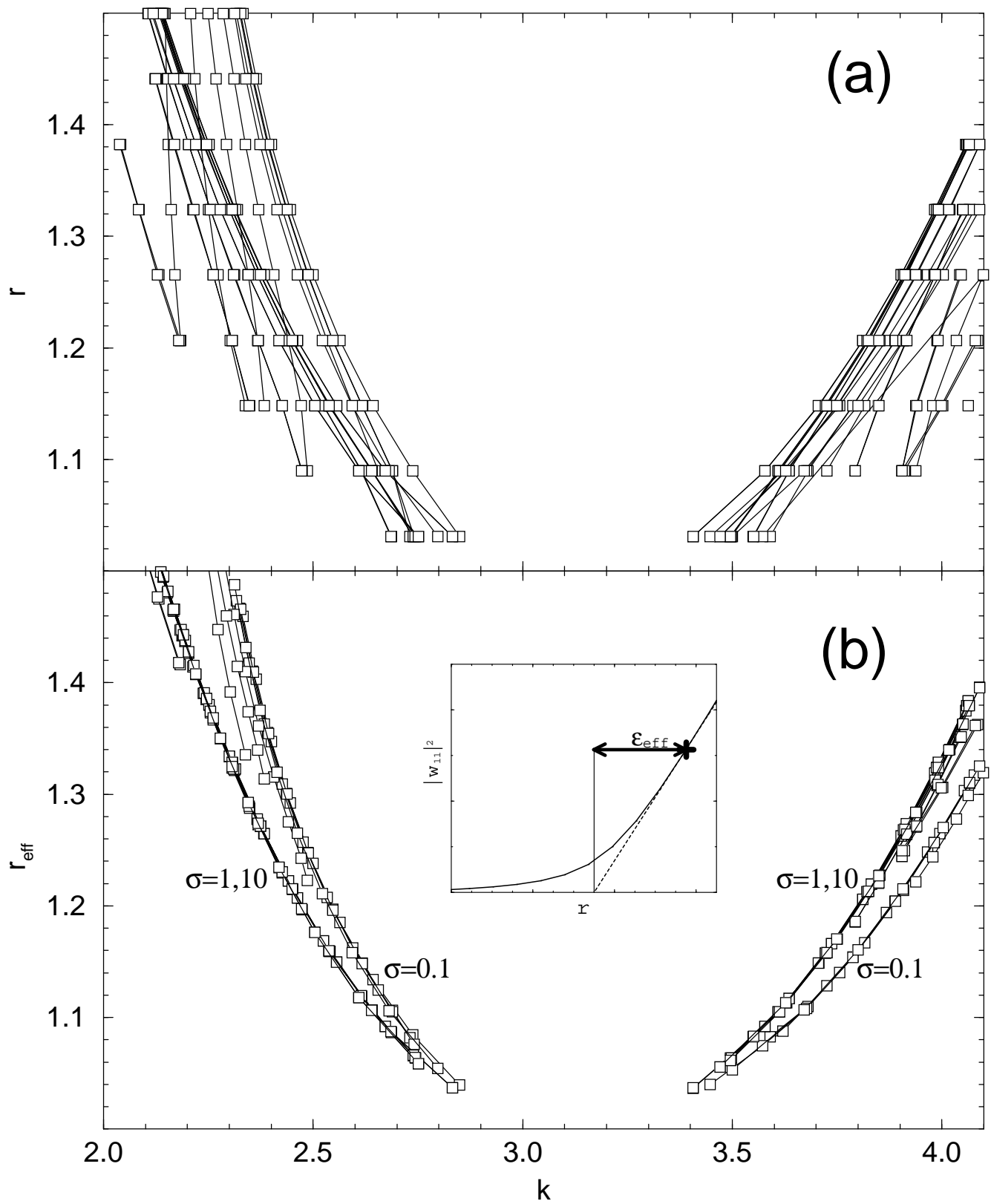
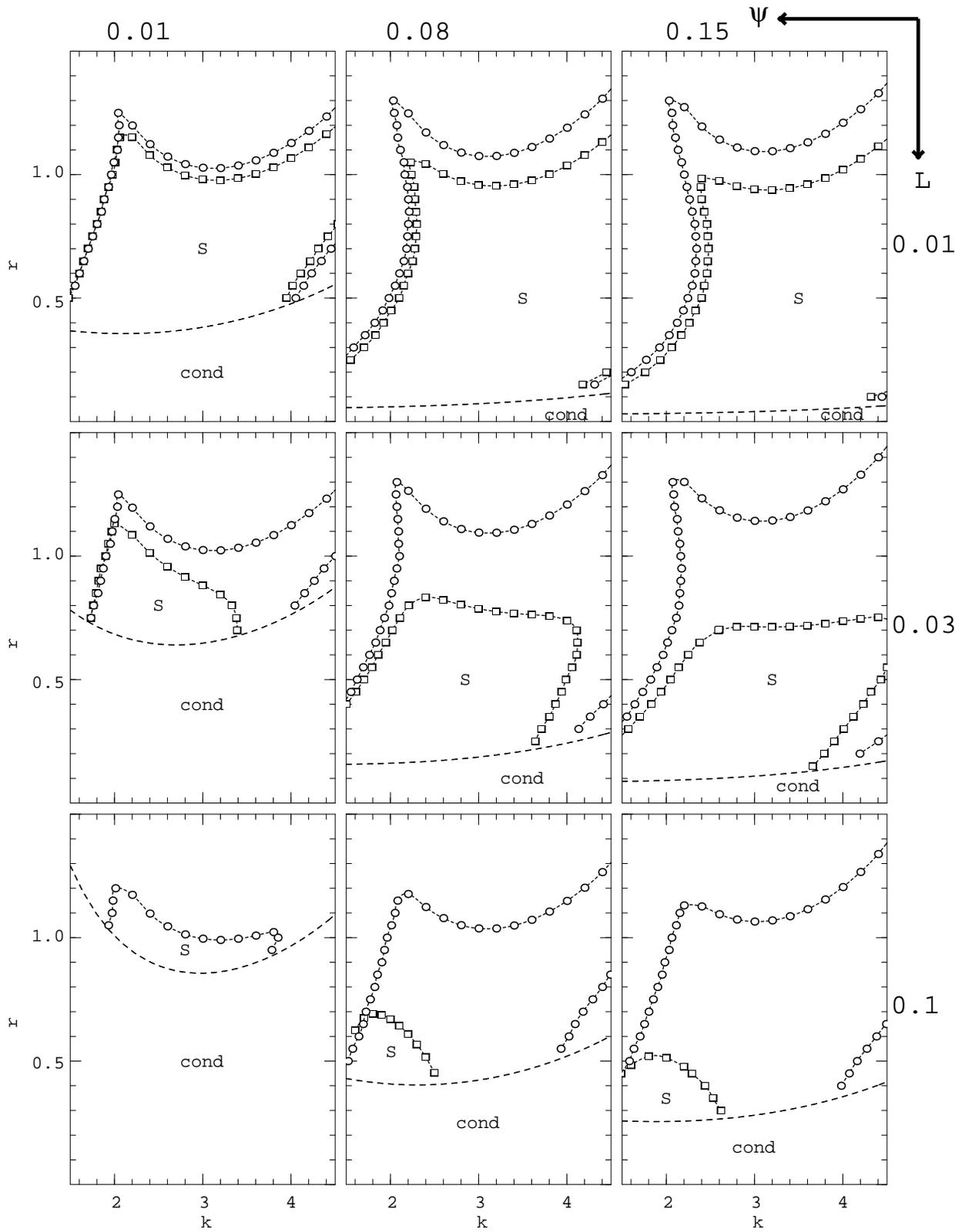
\end{document}